\documentclass[twocolumn,superscriptaddress,showpacs,prd,aps,amsmath,amssymb,nofootinbib]{revtex4-1}
\usepackage{graphicx}
\usepackage{amssymb}
\usepackage{bm}
\usepackage{color}
\usepackage{soul}

\graphicspath{{./}}

\newcommand{\Msol}{M_\odot}

\newcommand{\beq}{\begin{equation}} 
\newcommand{\eeq}{\end{equation}} 
\newcommand{\beqn}{\begin{eqnarray}} 
\newcommand{\eeqn}{\end{eqnarray}} 
\newcommand{\pa}{\partial}
\newcommand{\na}{\nabla}
\newcommand{\gab}{g^\alpha\!_\beta}
\newcommand{\gabu}{g^{\alpha\beta}}

\newcommand{\gmabd}{\gamma_{ab}}

\newcommand{\tgmabd}{\tilde\gamma_{ab}}
\newcommand{\tgamma}{\tilde\gamma}

\newcommand{\albe}{{\alpha\beta}}

\newcommand{\Tabu}{T^{\alpha\beta}}

\newcommand{\zD}{{\raise1.0ex\hbox{${}^{\ \circ}$}}\!\!\!\!\!D}
\newcommand{\alone}{{\raise0.5ex\hbox{${}^{\ 1}$}}\!\!\!\!\alpha}

\newcommand{\Dl}{\Delta}

\newcommand{\nalam}{\mathrel{\raise0.9ex\hbox{$^\lambda$}\mkern-14mu
\lower0.0ex\hbox{$\nabla$}}}

\newcommand{\gmaa}{\gamma^\alpha{}_a}
\newcommand{\gmbb}{\gamma^\beta{}_b}

\newcommand{\Nrf}{{N_r^{\rm f}}}
\newcommand{\Nrm}{{N_r^{\rm m}}}

\newcommand{\Kabd}{K_{ab}}

\newcommand{\zeroD}{{\raise1.0ex\hbox{${}^{\ \circ}$}}\!\!\!\!\!D}

\newcommand{\zLap}{{\raise1.0ex\hbox{${}^{\ \circ}$}}\!\!\!\!\Delta}
\newcommand{\zna}{{\raise1.0ex\hbox{${}^{\ \circ}$}}\!\!\!\!\!\nabla}
\newcommand{\zS}{{\raise1.0ex\hbox{${}^{\ \circ}$}}\!\!\!\!\!S}

\newcommand{\cocal}{\textsc{cocal}}

\newcommand{\GA}{\alpha}
\newcommand{\GB}{\beta}
\newcommand{\GG}{\gamma}
\newcommand{\GD}{\delta}
\newcommand{\GE}{\epsilon}
\newcommand{\GK}{\kappa}
\newcommand{\GL}{\lambda}
\newcommand{\GR}{\rho}

\newcommand{\GC}{\psi}

\newcommand{\GO}{\omega}

\newcommand{\GP}{\phi}

\newcommand{\Bn}{\mathbf{n}}

\newcommand{\Bt}{\mathbf{t}}

\newcommand{\BGB}{\boldsymbol{\GB}}
\newcommand{\Bphi}{\boldsymbol{\phi}}
\newcommand{\pd}{\partial}

\newcommand{\be}{\begin{equation}}
\newcommand{\ee}{\end{equation}}

\def\QEQ{{%
    \setbox0\hbox{$I$}%
    \rlap{\hbox to \wd0{\hss--\hss}}\box0
}}

\usepackage[]{amsmath}
\usepackage[]{dsfont}
\usepackage[]{amsfonts}
\usepackage[]{amssymb}
\usepackage[]{mathrsfs}
\usepackage{times}
\usepackage{hyperref}
\hypersetup{
  colorlinks=true,        
  linkcolor=blue,         
  citecolor=cyan,         
}
\newcommand{\re}[1]{{\textcolor{black}{#1}}}

\begin{document}

\title{Constant circulation sequences of binary neutron stars and their spin characterization}

\author{Antonios Tsokaros}
\affiliation{Department of Physics, University of Illinois at Urbana-Champaign,
Urbana, Illinois 61801}

\author{K\=oji Ury\=u}
\affiliation{Department of Physics, University of the Ryukyus, Senbaru, Nishihara, Okinawa 903-0213, Japan}

\author{Milton Ruiz}
\affiliation{Department of Physics, University of Illinois at Urbana-Champaign,
Urbana, Illinois 61801}
\author{Stuart L. Shapiro}
\affiliation{Department of Physics, University of Illinois at Urbana-Champaign,
Urbana, Illinois 61801}
\affiliation{Department of Astronomy \& NCSA, University of Illinois at
Urbana-Champaign, Urbana, Illinois 61801}

\date{\today}

\begin{abstract}
For isentropic fluids, dynamical evolution of a binary system conserves the
baryonic mass and circulation; therefore, sequences of constant rest mass and
constant circulation are of particular importance. In this work, we present the
extension of our Compact Object CALculator (\cocal{}) code to compute such
quasiequilibria and compare them with the well-known corotating and irrotational
sequences, the latter being the simplest, zero-circulation case.  The
circulation as a measure of the spin for a neutron star in a binary system  has
the advantage of being exactly calculable since it is a local quantity.  To
assess the different measures of spin, such as the angular velocity of the star,
the quasilocal, dimensionless spin parameter $J/M^2$, or the circulation
$\mathcal{C}$, we first compute sequences of single, uniformly rotating stars
and describe how the different spin diagnostics are related to each other. The
connection to spinning binary systems is accomplished through the concept of
circulation and the use of the constant rotational velocity formulation.
Finally, we explore a modification of the latter formulation that naturally
leads to differentially rotating binary systems. 
\end{abstract}

\maketitle

\section{Introduction}
\label{sec:intro}

Some of the most important problems in modern astrophysics include: a) the
origin of the heavy elements in the periodic table (heavier than iron), b) the
behavior of matter at densities beyond the nuclear, and c) the mechanism behind
the powerful electromagnetic events known as gamma-ray bursts, which in a few
seconds release as much energy as the sun does throughout its entire life. The
extreme conditions necessary for the creation of these phenomena can be found in
a binary neutron star (BNS) system through the combination of immense gravity,
electromagnetic fields, and nuclear forces. The 2017 detection of GW170817
confirmed these hypotheses and marked the birth of ``multimessenger astronomy''
since for the first time gravitational waves from a BNS system were directly
measured by the LIGO/VIRGO detector \cite{2017PhRvL.119p1101A} together with a
short duration gamma-ray burst by the Fermi Gamma-Ray Burst Monitor
\cite{2041-8205-848-2-L14} and INTEGRAL \cite{Savchenko:2017ffs}.

One of the most important characteristics of a neutron star (NS) is its
rotational frequency, which in isolation has been observed to be as high as
$716\;{\rm Hz}$, corresponding to a period of $1.4\;{\rm ms}$ for PSR
J1748-2446ad \cite{2006Sci...311.1901H}. In the 18 BNS systems currently known
in the Galaxy \cite{Tauris:2017omb,Zhu:2017znf}, the rotational frequencies are
typically smaller. The NS in the system J1807-2500B has a period of $4.2\;{\rm
ms}$, while systems J1946+2052 \cite{Stovall:2018ouw}, and J1757-1854
\cite{Cameron:2017ody}, J0737-3039A \cite{Kramer:2006nb} have periods $16.96,\
21.50,$ and $22.70\;{\rm ms}$, respectively.

Any evolution simulation of a BNS starts from initial data that describe the
system under consideration. The first such binary initial data were calculated
by Baumgarte \textit{et al.} \cite{1997PhRvL..79.1182B,1998PhRvD..57.7299B} and
Marronetti \textit{et al.} \cite{1998PhRvD..58j7503M} and described two NSs
tidally locked, as for example the Earth-Moon system. These were the so-called
corotating solutions, and although they gave the first insight into the problem,
they were rendered unrealistic since the viscosity is too small in NSs to
achieve synchronization \cite{1992ApJ...400..175B,1992ApJ...398..234K}.  A more
realistic scenario is the so-called irrotational state where the two NSs have
zero vorticity. Such systems were more difficult to describe and required an
additional potential equation. Irrotational BNS systems using different
numerical methods were presented by Bonazzola \textit{et al.}
\cite{1999PhRvL..82..892B}, Gourgoulhon \textit{et al.}
\cite{2001PhRvD..63f4029G}, Marronetti \textit{et al.}
\cite{2000NuPhS..80C0714M,1999PhRvD..60h7301M}, and Ury\=u \textit{et al.}
\cite{2000PhRvD..61l4023U,2000PhRvD..62j4015U}.  Even today, the majority of the
BNS simulations adopt these methods and therefore assume that the spin of the
individual NSs is zero. Such an assumption, although adequate in most cases,
cannot for example describe systems J1946+2052, J1757-1854, and J0737-3039A
which, according to Ref. \cite{Zhu:2017znf}, will have periods at merger of
$18.23,\ 27.09$, and $27.17\;{\rm ms}$, respectively. For accurate gravitational
wave analysis, one cannot consider these binaries to be irrotational, and the
spin of each NS must be taken into account.  Also, event GW170817
\cite{2017PhRvL.119p1101A} was unable to rule out high spin priors and thus two
sets of data (for low and high spins) were consistent with the observations.

Going beyond the two extreme cases of corotating and irrotational BNSs and
constructing binaries with arbitrary spin has proven to be more difficult due to
the fact that the Euler equation does not yield a trivial integral.  The first
attempt to address that problem was by Marronetti and Shapiro
\cite{Marronetti:2003gk}, who used instead the Bernoulli equation (first
integral along \textit{flow lines} and not globally) to construct sequences of
constant circulation. In Refs. \cite{2009PhRvD..80f4009B,2009PhRvD..80h9901B},
Baumgarte and Shapiro presented an alternative formulation to compute arbitrary
spinning binaries by constructing a new elliptic equation from the divergence of
the Euler equation. Although no solutions were presented there, violations of
the Euler equations were expected since their rotational part was not required
to vanish.  The only self-consistent formulation to obtain BNSs with arbitrary
spinning initial data was presented by Tichy \cite{Tichy:2011gw}, and
quasiequilibrium sequences were computed in Ref. \cite{2012PhRvD..86f4024T}. In
these studies, a first integral of the fluid flow was obtained under suitable
assumptions, and binary sequences with approximately constant rotational
velocity of each component were calculated. From a different perspective,
Tsatsin and Marronetti \cite{2013PhRvD..88f4060T} presented a method to produce
initial data for spinning BNSs that allowed for arbitrary orbital and radial
velocities, but without satisfying the Hamiltonian and momentum constraints.

In this work, we present the extension of our Compact Object CALculator
(\cocal{}) code for BNSs \cite{Uryu:2011ky,Tsokaros:2015fea,Tsokaros:2016eik} to
compute quasiequilibrium binary sequences of constant rest mass and circulation.
For isentropic fluids, dynamical evolution of a binary system conserves the
baryonic mass and circulation; therefore, sequences that conserve these
quantities can be considered realistic ``snapshots'' of an evolutionary
scenario.  We use Tichy's spinning formulation \cite{Tichy:2011gw} as we did in
Ref. \cite{Tsokaros:2015fea}, where sequences of constant rest mass alone were
computed, but focus here on the different spin measures that are currently used
\cite{Bernuzzi:2013rza,Dietrich:2015pxa,Tacik:2015tja} in order to make a
critical assessment. Using the circulation and rest mass as fundamental
properties, a connection between spinning companions in binaries and single
axisymmetric stars is established, and differences are discussed. Finally, we
present a decomposition alternative to Ref. \cite{Tichy:2011gw}, which slightly
simplifies the equations to be solved and leads naturally to differentially
rotating binary systems. Binary sequences of that kind are computed and compared
with the ones coming from the original formulation \cite{Tichy:2011gw}.

In this paper, spacetime indices are greek, spatial indices are latin, and the
metric signature is $-+++$.  For writing the basic equations, geometric units
with $G=c=1$ are used, while in all numerical solutions, $G=c=\Msol=1$ units are
used for convenience.

\section{Equations and general assumptions}

According to the first law of thermodynamics for binary systems by Friedman
\textit{et al.} \cite{Friedman:2001pf,Friedman:2013xza}, if one assumes a
spatial geometry $\Sigma_t$ that is conformally flat, neighboring equilibria of
asymptotically flat spacetimes with a helical Killing vector 
asymptotic form $k^\GA=t^\GA+\Omega\GP^\GA$ ($t^\GA$ and satisfy 
\begin{eqnarray} \GD M & = & \Omega \delta J + \int_{\Sigma_t}
   [\bar{T}\Delta dS+\bar{\mu}\Delta dM_B+V^\GA\Delta dC_\GA] \nonumber \\ & + &
\sum_i \frac{1}{8\pi}\GK_i \GD A_i .  \label{eq:1stlaw} 
\end{eqnarray} 
Here, $M$ and $J$ are the Arnowitt-Deser-Misner (ADM) mass and angular momentum
of the spacetime, while $\Omega$ is the orbital angular velocity; $\bar{T}$ and
$\bar{\mu}$ are the redshifted temperature and chemical potential; $dM_B$ is the
baryon mass of a fluid element; $dC_\GA$ is related to the circulation of a
fluid element and $V^\GA$ is the velocity with respect to the corotating frame;
and $\GK_i$ and $ A_i$ are the surface gravity and the areas of black holes.
For isentropic fluids, dynamical evolution conserves the baryon mass, entropy,
and vorticity of each fluid element, and thus the first law yields $ \GD M =
\Omega \delta J $. Equation (\ref{eq:1stlaw}) implies that a natural measure to
characterize the spin of a NS in a binary setting is its circulation in a manner
similar to the way rest mass characterizes the mass. Since different spin
measures are used in BNS studies
\cite{Bernuzzi:2013rza,Dietrich:2015pxa,Tacik:2015tja}, one question that arises
is how all these diagnostics are related to the conserved quantity of
circulation. Before answering this question we will investigate the relationship
of these quantities for single, axisymmetric, rotating stars.  We will adopt the
3+1 formulation of \cite{Uryu:2016dqr} in order to make contact with the theory
of a single rotating star, while for BNS systems, we will use the notation of
Ref.  \cite{Tsokaros:2015fea}. The equations solved are reported in detail in
those two papers, so here we will only review the necessary definitions and
assumptions in a unified way.

We assume that the spacetime $\cal M$ is asymptotically flat and is foliated by
a family of spacelike hypersurfaces $(\Sigma_t)_{t\in {\mathbb R}}$,
parametrized by a time coordinate $t\in {\mathbb R}$ as ${\cal M} = {\mathbb R}
\times \Sigma_t$ \cite{BSBook}. The future-pointing unit normal one form to
$\Sigma_t$, $n_\alpha := -\alpha\na_\alpha t$, is related to the generator of
time translations $t^\alpha$ as $t^\alpha := \alpha n^\alpha + \beta^\alpha$,
where $t^\alpha \na_\alpha t = 1$. $\alpha$ and $\beta^\alpha$ are,
respectively, the lapse and shift and $\beta^\alpha$ is spatial, $\beta^\alpha
\na_\alpha t=0$. The projection tensor to $\Sigma_t$ $\gamma_{\alpha}{}^{\beta}$
is introduced as $\gamma^{\alpha}{}_{\beta} := \gab+n^\alpha n_\beta$. The
induced spatial metric $\gmabd$ on $\Sigma_t$ is the projection tensor
restricted to it. Introducing a conformal factor $\psi$, and a conformally
rescaled spatial metric $\tgmabd$, the line element on a chart $\{t,x^i\}$ of
$\Sigma_t$ is written as
\beq
ds^{2}=-\alpha^{2}dt^{2}+\psi^{4}\tgamma_{ij} (dx^i+\beta^i dt)(dx^j+\beta^j dt).
\eeq
The conformal rescaling is determined from a condition $\tgamma =f$, where
$\tgamma$ and $f$ are determinants of the rescaled spatial metric $\tgmabd$ and
the flat metric $f_{ab}$.  In what follows, we will assume that
$\tgamma_{ij}=f_{ij}$ for both single and binary star computations.

The extrinsic curvature of each slice $\Sigma_t$ is defined by 
\beqn
\Kabd &:=& -\frac12 \gmaa \gmbb \mathcal{L}_{\Bn} \gamma_{\albe},  \nonumber \\
&=& -\frac1{2\alpha}\pa_t \gmabd + \frac1{2\alpha}\mathcal{L}_{\BGB} \gmabd ,
\label{eq:Kab}
\eeqn
where $\pa_t \gmabd$ is the pullback of $\mathcal{L}_{\Bt} \gamma_\albe$ to
$\Sigma_t$, $\mathcal{L}_{\Bt}$ is the Lie derivative along the vector
$t^\alpha$ defined on $\cal M$, and $\mathcal{L}_{\BGB}$ is the Lie derivative
along the spatial vector $\beta^a$ on $\Sigma_t$.  Hereafter, we denote the
trace of $K_{ab}$ by $K$, and the trace-free part of $K_{ab}$ by $A_{ab} :=
K_{ab} -\frac13\gmabd K $.  For both single and BNS systems we will assume the
maximal slicing condition
\be
K=0 .
\ee

In this paper, we consider perfect-fluid spacetimes in which the stress-energy
tensor is written as \cite{Tsokaros:2015fea}
\beq
\Tabu := (\epsilon+p)u^\alpha u^\beta + p\gabu, 
\label{eq:Tab}
\eeq
where $\epsilon$ is the energy density, $p$ is the pressure, and $u^\alpha$ is
the 4-velocity.  The relativistic enthalpy $h$ is defined as 
\be
h :=\frac{\GE+p}{\GR}, 
\label{eq:h}
\ee
where $\GR$ is the rest mass density. The 4-velocity of the fluid can be written
as $u^\GA=u^t(1,v^i)$ and, in analogy to a Newtonian decomposition, we can split
the spatial component $v^i$ into two parts: one that follows the rotation around
the center of mass, $\Omega\GP^i$, and one that represents the velocity in the
corotating frame $V^i$, 
\be
u^\GA := u^t (t^\GA + v^\GA) = u^t (k^\GA + V^\GA), 
\label{eq:4vel}
\ee
where $v^\GA=(0,v^i) :=\Omega\GP^\GA+V^\GA$, and 
\be
k^\GA := t^\GA + \Omega\GP^\GA = \GA n^\GA + \GO^\GA\, .
\label{eq:hkv}
\ee
Here, the helical Killing vector $k^\GA$ applies to either a binary system
having orbital angular velocity $\Omega$ or a single rotating star
(axisymmetric or not) having the same  constant, rotating angular velocity. The
vector $\GO^\GA := \GB^\GA+\Omega\GP^\GA$ is the so-called corotating shift.
For single rotating stars as well as for corotating binaries, $V^\GA=0$. 

Fluid variables will be computed through the conservation of the energy-momentum
tensor
\begin{eqnarray}
0 & = & \nabla_\GA T^{\GA}_{\ \; \GB}  \nonumber \\
  & = & \GR[u^\GA\nabla_\GA(hu_\GB)+\nabla_\GB h - T\nabla_\GB s] + hu_\GB\nabla_\GA(\GR u^\GA)  \nonumber \\
  & = & \GR[u^\GA\GO_{\GA\GB}-T\nabla_\GB s] + hu_\GB\nabla_\GA(\GR u^\GA) \, ,
\end{eqnarray}
and local conservation of rest mass
\begin{equation}
\nabla_\GA(\GR u^\GA)=0\, .  \label{eq:cbm}
\end{equation}
Assuming isentropic configurations, the relativistic Euler equation becomes
$u^\GA\GO_{\GA\GB}=0$, where
\begin{equation}
\GO_{\GA\GB}:=\nabla_{\GA}(hu_\GB) - \nabla_{\GB}(h u_\GA)\,   \label{eq:rvt}
\end{equation}
is the relativistic vorticity tensor, which is zero for irrotational flow
\cite{Rezzolla_book:2013}.

In 3+1 language, the Euler equation and the rest mass conservation equation
become \cite{Tsokaros:2015fea}
\begin{eqnarray}
\GG_i^\GA \mathcal{L}_{\boldsymbol k}(hu_\GA) + D_i\left(\frac{h}{u^t} + hu_j V^j\right) + V^j\GO_{ji} = 0\, , \label{eq:ree3}  \\
\mathcal{L}_{\boldsymbol k}(\GR u^t) + \frac{1}{\GA}D_i(\GA\GR u^t V^i) = 0\, , \quad\qquad\qquad   \label{eq:cbm3}
\end{eqnarray}
where $D$ is the covariant derivative with respect to the spatial metric,
$D_a\GG_{ij}=0$.

For single rotating stars, as well as for corotating binaries under the helical
symmetry assumption, Eq. (\ref{eq:cbm3}) is trivially satisfied, while the Euler 
equation results in a simple algebraic equation,
\begin{equation}
\frac{h}{u^t} = C\,,  \label{eq:corotei}
\end{equation}
where $C$ is a constant to be determined and $u^t = 1/\sqrt{\GA^2 - \GO_i
\GO^i}$.

For irrotational binaries \cite{1997PhRvD..56.7740B,1998PhRvD..57.7292A,
Shibata:1998um, Teukolsky:1998sh}, we have $\GO_{\GA\GB}=0$, so the specific
enthalpy current $hu_\GA$ can be derived from a potential $hu_\GA = \nabla_\GA
\Phi$. In order to allow for arbitrary spinning binary configurations, a
3-vector $s^i$ is introduced according to \cite{Tichy:2011gw}
\begin{equation}
\hat{u}_i := \GG_i^\GA hu_\GA = D_i\Phi + s_i\,,  \label{eq:uih_dec}
\end{equation}
where the $D_i\Phi$ part corresponds to the ``irrotational part'' of the flow
and $s^i$ the ``spinning part'' of the flow.  In our code vector $s^i$ is the
input quantity, and $s_i = \GG_{ij}s^j$.  For a general vector $s^i$, one can
have a binary system that exhibits differential rotation.  Irrotational binaries
are recovered for $s^i=0$.  According to Ref. \cite{2012PhRvD..86f4024T} a
choice that minimizes differential rotation is a rigid rotation law,  
\be
s^{i} := \Omega_{\rm s}^a \GP^{i}_{{\rm s}(a)}   \label{eq:spin}
\ee
where $\GP^{i(a)}_{\rm s}=\GE^{iaj}X_j$ denotes the rotation vectors along the
NS's three axes.  \re{The index $i$ corresponds to the component of the vector
$\Bphi^{(a)}_{\rm s}$, while the index inside the parenthesis names the three
different vectors.  Vector $\GP^{i(3)}_{\rm s}$, which in the following sections
is denoted by $\GP^i_{\rm s}$, is the rotation vector along the star's
\re{$X_3$-axis}, in contrast to $\GP^i$ which is the rotation vector along the
z-axis. For single rotating stars these two vectors are identical.  We denote by
$x_i=\{x,y,z\}$ the coordinates around the center of mass of the binary system,
and by $X_a=\{X_1,X_2,X_3\}$ the coordinates centered at the maximum density
point of each NS.  The orbital vector $\GP^i$ refers to $\{x,y,z\}$ while the
spin vector $s^i$ refers to $\{X_1,X_2,X_3\}$. In this work we assume that the
rotation of the neutron stars is around $X_3$. The z-axis and the $X_3$ axis are
parallel and perpendicular to the orbital plane.  The coefficients $\Omega_{\rm
s}^a$ are parameters that control the rotational spin around the NS's three axes
$X_a$}.  These parameters, although lacking of physical (i.e. invariant)
meaning, approximately represent the angular velocity of the rotating star.

From Eqs. (\ref{eq:4vel}) and (\ref{eq:uih_dec}), the spatial velocity $V^i$ of
the flow is
\begin{equation}
V^i=\frac{D^i\Phi+s^i}{hu^t}-\GO^i\,. \label{eq:irspuih}
\end{equation}

For arbitrary spinning binaries, the Euler equation (\ref{eq:ree3}) becomes
\begin{equation}
\GG_i^\GA \left[ \mathcal{L}_{\boldsymbol k}(hu_\GA) + 
\mathcal{L}_{\boldsymbol V}(s_\GA) \right] + 
D_i\left(\frac{h}{u^t} + V^j D_j\Phi\right) = 0\,, 
\end{equation}
which under the assumptions of helical symmetry and the additional assumption of
\be
\mathcal{L}_{\boldsymbol V}(s_\GA) = 0  \label{eq:assumspin}
\ee
yields
\begin{equation}
\frac{h}{u^t} + V^j D_j\Phi = C\,, 
\label{eq:irspei}
\end{equation}
where again $C$ is a constant to be determined. Although the Euler integral has
the same form for both irrotational and spinning binaries, it produces a
different equation since the 3-velocity $V^i$ is different in these two cases.
Assumption (\ref{eq:assumspin}) means that changes of the spin vector with
respect to the corotating velocity are small.

The normalization condition $u_\GA u^\GA=-1$, together with Eqs.
(\ref{eq:uih_dec}), (\ref{eq:irspuih}), and (\ref{eq:irspei}), yield
\begin{eqnarray}
hu^t & = & \frac{\GL+\sqrt{\GL^2+4\GA^2 s_i(D^i\Phi + s^i)}}{2\GA^2}\,,   \label{eq:hutsol} \\
h & = & \sqrt{\GA^2(hu^t)^2 - (D_i\Phi+s_i)(D^i\Phi+s^i)}  .   \label{eq:hsol}
\end{eqnarray}
Here, $\GL :=C+\GO^i D_i\Phi$.  For purely irrotational binaries,
$hu^t=\GL/\GA^2$
and $h=\sqrt{\GL^2/\GA^2 - D_i\Phi D^i\Phi}$.  The fluid potential $\Phi$ is
computed from the conservation of rest mass (\ref{eq:cbm3}) and the use of Eqs.
(\ref{eq:irspei}) and (\ref{eq:irspuih}) \cite{Tsokaros:2015fea}.

\begin{figure}
\begin{center}
\includegraphics[width=0.99\columnwidth]{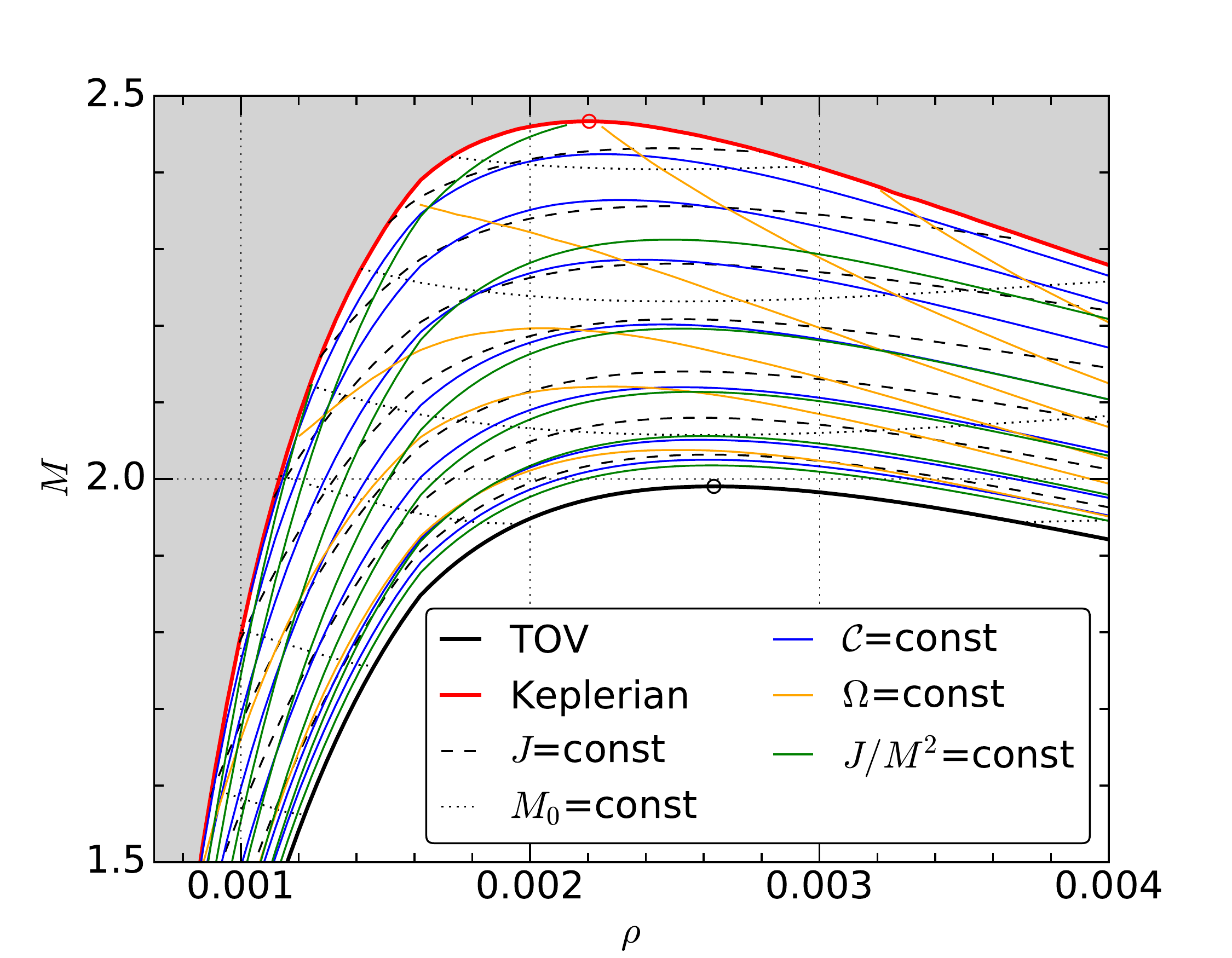}
\caption{Mass vs rest-mass density for sequences of uniformly rotating single
stars with constant angular momentum $J$, rest mass $M_0$, circulation
$\mathcal{C}$, angular velocity $\Omega$, dimensionless spin $J/M^2$, together
with the spherical (TOV) and mass-shedding (Kepler) limits.}  
\label{fig:ALF2star}
\end{center}
\end{figure}

\section{Measures of spin and constant circulation sequences}

\subsection{Single stars}
For single rotating stars, one has a variety of ways to characterize the spin.
Among them are its angular velocity $\Omega$ (we assume constant rotation), its
ADM angular momentum $J$ 
\be
J = \frac{1}{8\pi} \int_{S_\infty} K^a_{\ \, b}\GP^b dS_a
\label{eq:angmom}
\ee
or the dimensionless spin $J/M^2$, where $M$ is the ADM mass.
Using Gauss's theorem, Eq. (\ref{eq:angmom}) can be written as
\begin{eqnarray}
J & = & \frac{1}{8\pi}\int_{V_t} D_a(K^a_{\ \, b} \GP^b) d\Sigma - 
        \frac{1}{8\pi} \int_S K^a_{\ \, b}\GP^b dS_a   \nonumber \\
  & = & \frac{1}{8\pi}\int_{V_t} K^a_{\ \, b} \pd_a \GP^b d\Sigma - 
      \frac{1}{8\pi} \int_S K^a_{\ \, b}\GP^b dS_a   \label{eq:Jvol}
\end{eqnarray}
where $\pd V_t=S_\infty \cup S$. To go from the first volume integral to the
second, we used the maximal slicing assumption and the momentum constraint with
zero sources since $S$ is taken to be outside the fluid volume. Without loss of
generality, we can assume the $S$ is a sphere just outside the surface of the NS.

When the conformal geometry is flat (as happens in most binary neutron star
calculations), $\GP^a$ is a Killing vector of the conformal geometry, and
therefore the volume integral in Eq. (\ref{eq:Jvol}) is zero. We call the remain
integral the quasilocal spin angular momentum 
\be
J_{\rm ql} = \frac{1}{8\pi} \int_S K^a_{\ \, b}\GP^b dS_a
\label{eq:qlspin}
\ee
where here the unit normal is outward. Thus, under the assumptions of conformal
flat geometry and maximal slicing, 
\be
J=J_{\rm ql}\qquad\mbox{(single stars)} \ .
\label{eq:jeqjqls}
\ee

Another way to measure the spin of a rotating star is by its circulation. For
rotation around the z axis, 
\begin{eqnarray}
\mathcal{C} & := & \oint_c hu_\GA dx^\GA = \oint_c hu^t \GC^4 \GD_{ij}(\GB^i + \Omega\GP^i)dx^j \ , \label{eq:circstar} 
\end{eqnarray}
where $c$ can be taken to be a fluid equatorial ring. One of the advantages of
using the circulation as a spin diagnostic is the fact that Eq.
(\ref{eq:circstar}) is local in character and involves quantities that are
exactly known (essentially the fluid velocity).  Although all single rotating
star models reported in this paper are axisymmetric, we have checked our
circulation code in the case of single triaxial stars
\cite{Uryu:2016dqr,Uryu:2016pto}, where the curve $c$ is no longer a circle but
close to an ellipse. 

In order to understand how the different measures of spin are related to each
other for single uniformly rotating stars, we use the \cocal{} code
\cite{Uryu:2016dqr} to build sequences of constant angular momentum $J$,
circulation $\mathcal{C}$, angular velocity $\Omega$, and dimensionless spin
$J/M^2$, together with the spherical (TOV) and mass-shedding (Kepler) limits as
in Fig. \ref{fig:ALF2star}.  For the equation of state (EoS), we have chosen the
piecewise representation of ALF2 \cite{2005ApJ...629..969A,Read:2008iy}, which
according to event GW170817 it is still a viable choice. Having said that, we
point out that the results of this work do not depend on this choice and any
other EoS would have been as good for conveying the ideas we put forward here.
In our code, we compute the circulation both as the line integral
(\ref{eq:circstar}) and also as a surface integral using Stokes's theorem. Both
quantities agree to the precision of our calculation, which is less than $1\%$.
The curve $c$ is chosen to be along the surface of the star in the $xy$ plane,
which, according to our normalization scheme (use of surface fitted
coordinates), is the unit circle \cite{Uryu:2016dqr}.  From the computational
point of view, one important aspect of the \cocal{} code is the use of
normalized coordinates for both single rotating stars \cite{Uryu:2016dqr} as
well as binaries \cite{Tsokaros:2015fea},
\be
\hat{x}^i := \frac{x^i}{R_0},\qquad \hat{\Omega} := \Omega R_0 \ .
\label{eq:nc}
\ee
The normalization factor that determines the length scale $R_0$ is only found at
the end of the iteration procedure and varies at every iteration. The constants
$R_0$, $\hat{\Omega}$, and $C$ [from the hydrostatic equilibrium
(\ref{eq:corotei}), (\ref{eq:irspei})] are determined through a solution of a
nonlinear $3\times 3$ system as described in Refs.
\cite{Uryu:2016dqr,Tsokaros:2015fea}. 

In terms of the normalized quantities,
\be
\mathcal{C} = R_0 \oint_c hu^t \GC^4 \GD_{ij}(\GB^i + \hat{\Omega}\hat{\GP}^i) d\hat{x}^j  \ .
\ee

\begin{figure}
\begin{center}
\includegraphics[width=0.99\columnwidth]{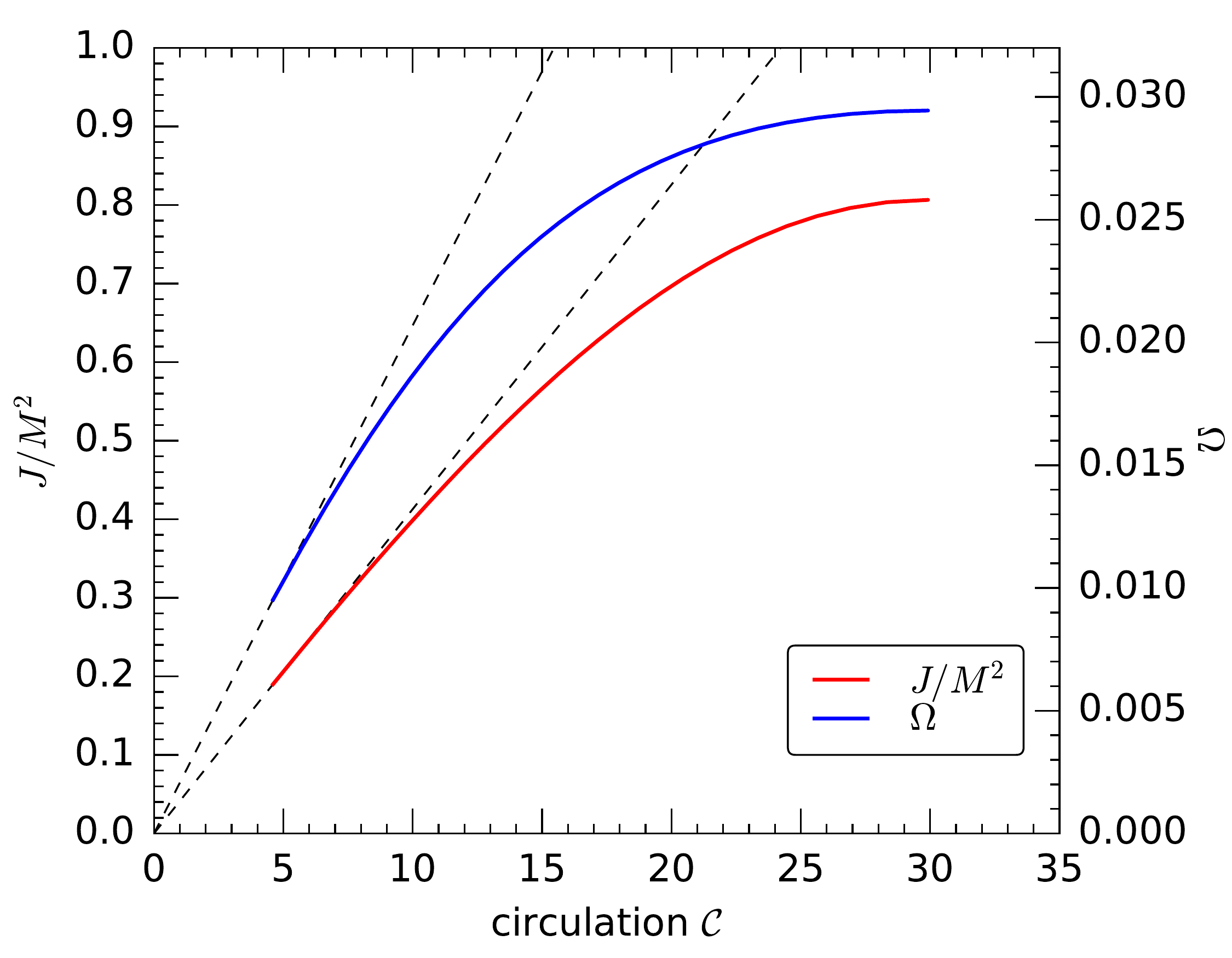}
\caption{Dimensionless spin parameter $J/M^2$ and angular velocity $\Omega$ vs
circulation $\mathcal{C}$ along a sequence of uniformly rotating, single stars
with constant rest mass $M_0=1.5$.}  
\label{fig:circ_jm2}
\end{center}
\end{figure}

As one can see from Fig. \ref{fig:ALF2star}, all curves that measure the spin of
a rotating star are \textit{in general distinct}. If a set of curves A is
``parallel'' to another set of curves B, this means that a star that is moving
along a constant A sequence will also move along a constant B sequence, or in
other words, conservation of quantity A will imply the conservation quantity B.
As far as the different spin measures ${J,\Omega,\mathcal{C},J/M^2}$ for
rotation close to the mass-shedding limit (red curve), this cannot happen since
all sets of curves are distinctly different. By contrast, close to the spherical
limit, one can see that constant circulation sequences are almost parallel to
constant $J/M^2$ sequences.  \re{This means that the curve $\mathcal{C}=c_1$
will nearly coincide with a curve $J/M^2=c_2$, (for two constants $c_1\neq c_2$)
when rotation is slow,} and therefore if during a process one parameter is
conserved, so is the other.  In Fig. \ref{fig:circ_jm2}, we plot the
dimensionless spin $J/M^2$ and angular velocity $\Omega$ vs the circulation for
a sequence of constant rest mass $M_0=1.5$. Dashed black lines connect the first
points of the sequences to the static limit (TOV).  Along that sequence, the ADM
mass varies approximately from $1.35$ to $1.39$.  As we can see for
dimensionless spins up to $\sim 0.4$, the two quantities vary linearly, but for
higher spins, especially close to the mass-shedding limit, this dependence
becomes quadratic. Beyond this point, increasing the circulation results in a
smaller increase in $J/M^2$.

\begin{table*}
\begin{tabular}{cl|ccccccccccccc}
\hline
\hline
Type & Patch  & $\ r_a\ $ & $\ r_s\ $ & $\ r_b\ $ & $\ r_c\ $ & $\ r_e\ $ & 
$\ \Nrf\ $ & $\ N_r^1\ $ & $\ \Nrm\ $ & $\ N_r\ $ & $\ N_\theta\ $ & $\ N_\phi\ $ & $\ L\ $  \\
\hline
$\texttt{Hd2.0}$ & ${\rm COCP-1}$ & $0.0$ & varies & $10^2$ & varies & $1.125$ 
               & $50$  & $64$ & $80$ & $192$ & $48$  & $48$ & $12$  \\
               & ${\rm COCP-2}$ & $0.0$ & varies & $10^2$ & varies & $1.125$ 
                                                         & $50$  & $64$ & $80$ & $192$ & $48$  & $48$ & $12$ \\
               & ${\rm ARCP}$   & $5.0$ & $-$         & $10^6$ & $6.25$ & $-$     
                                                         & $16$  & $-$  & $20$ & $192$ & $48$  & $48$ & $12$ \\
\hline
\hline
\end{tabular}
\caption{Grid structure parameters used for the binary computation in \cocal{}.
$r_a$ is the radial coordinate where the grids start, 
$r_b$ the radial coordinate where the grids end,
$r_c$ the center-of-mass point (excised sphere is located at $2r_c$), 
$r_e$ is the radius of the excised sphere, 
$r_s$ is the radius of the sphere bounding the star's surface, 
$N_{r}$ is the number of intervals $\Dl r_i$ in $r \in[r_a,r_{b}]$,
$N_{r}^{1}$ is the number of intervals $\Dl r_i$ in $r \in[0,1]$,
$\Nrf$ is the number of intervals $\Dl r_i$ in $r \in[0,r_s]$,
$\Nrm$ is the number of intervals $\Dl r_i$ in $r \in[r_a,r_{c}]$,
$N_{\theta}$ is the number of intervals $\Dl \theta_j$ in $\theta\in[0,\pi]$,
$N_{\phi}$ is the number of intervals $\Dl \phi_k$ in $\phi\in[0,2\pi]$,
and $L$ is the order of included multipoles.
 Distances are in normalized quantities, and $r_s$ varies during the iterations in order 
for a specific distance (angular velocity) to be reached.
For more details, see Refs. \cite{Tsokaros:2015fea, Uryu:2011ky}.}
\label{tab:cocgrids}
\end{table*}

\subsection{Binary stars}

\begin{figure}
\begin{center}
\includegraphics[width=0.99\columnwidth]{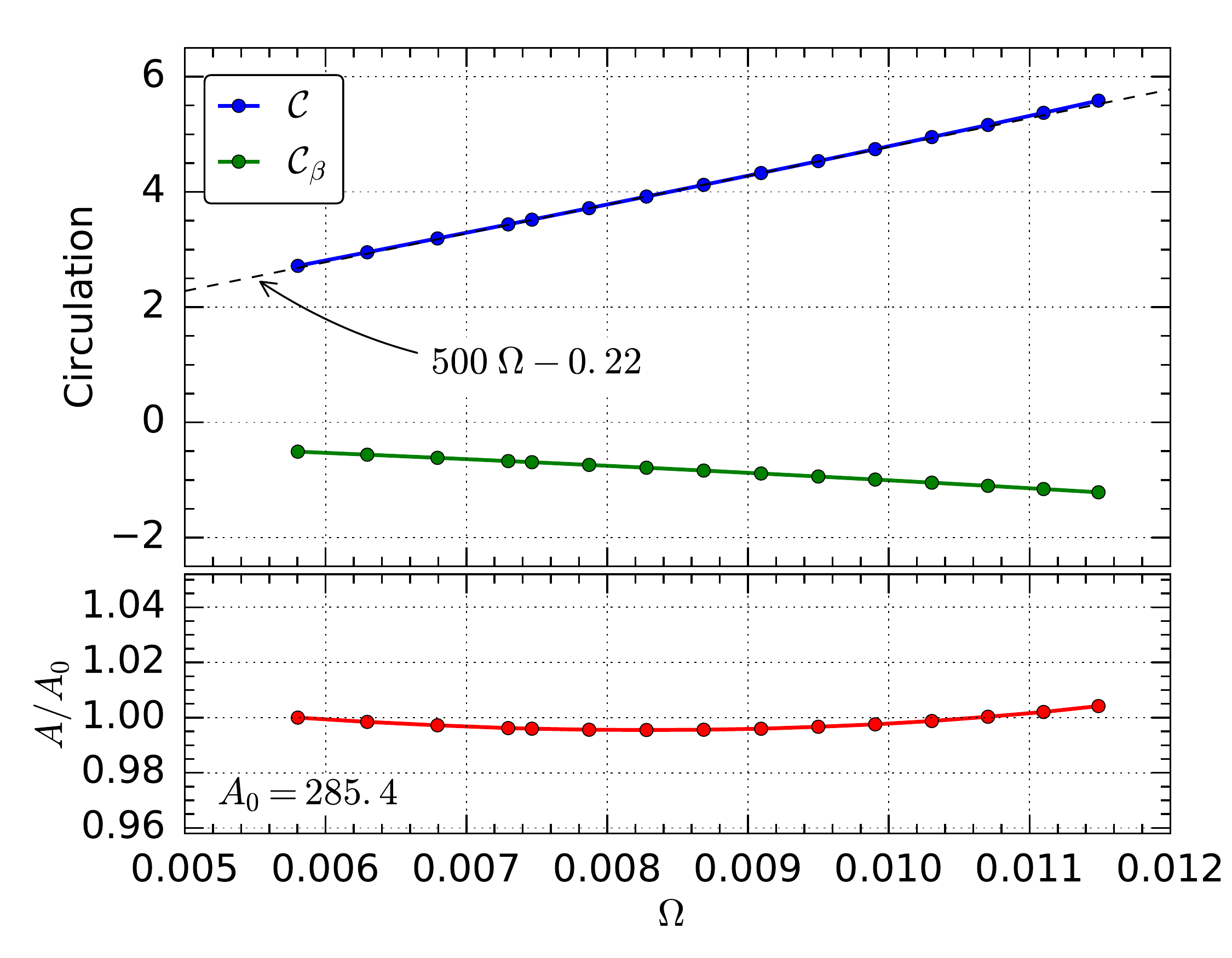}
\caption{The top panel shows circulation $\mathcal{C}$ and coordinate
circulation $\mathcal{C}_\GB$ for a corotating BNS sequence of constant rest
mass $M_0=1.5$. The bottom panel shows the approximate equatorial area $A\approx
2\pi R_x R_y$ of the each NS along the sequence. Values are normalized by $A_0$,
the area of the first member of the sequence.}  
\label{fig:cocirc}
\end{center}
\end{figure}

For a corotating binary, the circulation of each star is given by the same
formula as in a single rotating star (\ref{eq:circstar}) where now the vector
$\GP^i$ is the z-rotational vector (we assume the binary orbit to be in the
xy-plane) around the center of mass.  In Fig. \ref{fig:cocirc}'s top panel, we
plot the circulation $\mathcal{C}$ and the ``coordinate circulation''
$\mathcal{C}_\GB :=\oint_c hu^t \GC^4 \GD_{ij} \GB^i dx^j$ for a constant rest
mass sequence with $M_0=1.5$ as a function of $\Omega$.  In the bottom panel, we
plot the approximate coordinate equatorial area of each star $A\approx 2\pi R_x
R_y$, normalized by its initial value in the sequence $A_0$.  As we can see, the
circulation increases linearly with respect to the angular velocity, which
provides yet another argument as to why the corotating state is not realistic
for BNS systems with isentropic fluids. In the Newtonian limit,
$\mathcal{C}=2A\Omega$, where $A$ is the equatorial area of the NS. From the
bottom panel of Fig. \ref{fig:cocirc}, we see that the equatorial area is
approximately conserved along the sequence. Therefore, the circulation of the
corotating sequence follows essentially the Newtonian law apart from a redshift
factor. 
We also observe that, even when they are close to each
other, the circulation of the corotating binaries is relatively small compared
to the maximum circulation $\mathcal{C}_{\rm max}\approx 30$ for the ALF2 EoS
for single rotating stars. From Fig. \ref{fig:circ_jm2}, this implies
dimensionless spins lower than say $\sim 0.4$ (we calculate below the exact
values).  The coordinate circulation $\mathcal{C}_\GB$ (green curve) has
opposite sign from $\mathcal{C}$ and typically grows also linearly and is $\sim
20\%$ of $\mathcal{C}$.  For all binary calculations in this work, we used grid
values as reported in Table \ref{tab:cocgrids}.  \re{In order to create binaries
at different separations we choose $r_c\in \{1.125,1.25,1.50,1.75\}$ where
$1.125$ leads to close binaries, while $1.75$ to widely separated ones
\cite{Tsokaros:2015fea}.}

For an irrotational binary, the circulation is zero since the enthalpy current
$hu_\GA$ is a total derivative. For spinning binaries with 4-velocity
(\ref{eq:uih_dec}) and spin along the orbital axis, the circulation becomes
\be
\mathcal{C} = \oint_c s_i dx^i = R_0 \oint_c \GC^4 \GD_{ij}\hat{\Omega}_{\rm s}\hat{\GP}^i_{\rm s} d\hat{x}^i \, , 
\label{eq:circbin}
\ee
where code (normalized) coordinates (\ref{eq:nc}) are used.  Here, $\Omega_{\rm
s}^a=(0,0,\Omega_{\rm s})$ and $s^{i} := \Omega_{\rm s}^3 \GP^{i}_{{\rm s}(3)}
$.  Sequences of constant rest mass for fixed values of $\hat{\Omega}_s$ have
been calculated in Ref. \cite{Tsokaros:2015fea}. Here, we have extended our
\cocal{} code \cite{Uryu:2011ky,Tsokaros:2015fea} in order to compute binary
sequences of both constant circulation and rest mass.  In order to do that, a
multiroot secant method was implemented, which in principle can iterate over
different quantities like densities, spins, or distances in order to achieve
some target values. The computational cost, though, for such a finder increases
considerably.  In particular, the method converges after approximately ten
cycles and for each cycle, one needs $N_i$ converged solutions, where $N_i$ is
the number of quantities that we are targeting.  For equal-mass binaries that we
calculate here, in order to find a sequence of constant rest mass and
circulation, ($N_i=2$) $\sim 20$ converged solutions are needed.  If one also
insists these binary separations  are at a certain distance (or angular
velocity), then $N_i=3$. For each converged solution, one needs $\sim 500$
iterations.  Also in this work, we assume symmetric aligned or antialigned
binaries; i.e. we only have to search for one out of the six spin components.
For the general case, the computational cost will increase by an order of
magnitude.

\begin{figure}
\begin{center}
\includegraphics[width=0.99\columnwidth]{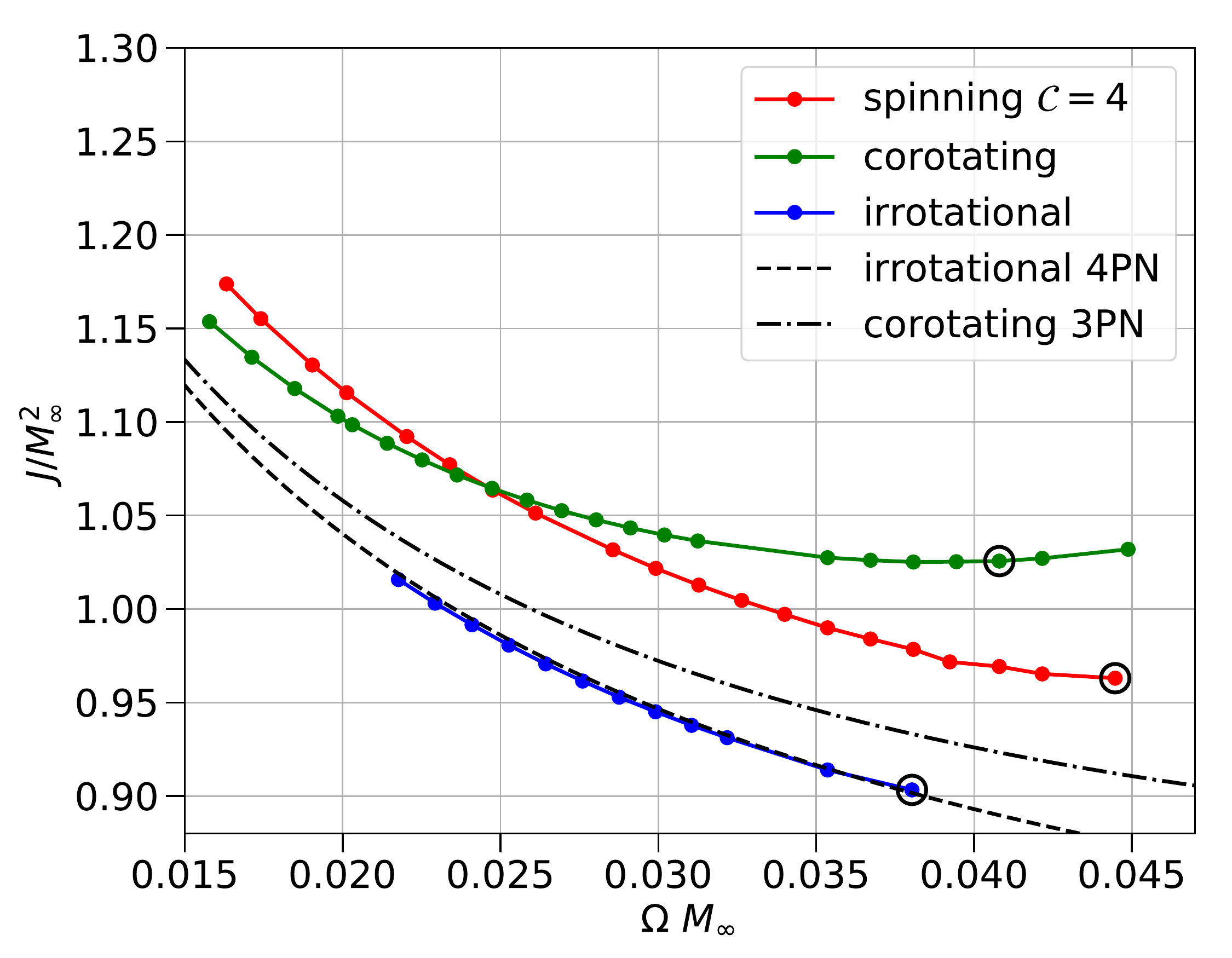}
\caption{Angular momentum curve for a binary sequence with constant circulation
$\mathcal{C}=4$ and rest mass $M_0=1.5$, along with the typical corotating and
irrotational sequences of the same rest mass. Points marked with a larger black
circle denote the approximate ISCO. Realistic physical sequences have constant
circulation and rest mass, such as the red or blue one.}  
\label{fig:ALF2binJM2}
\end{center}
\end{figure}

\begin{figure}
\begin{center}
\includegraphics[width=0.99\columnwidth]{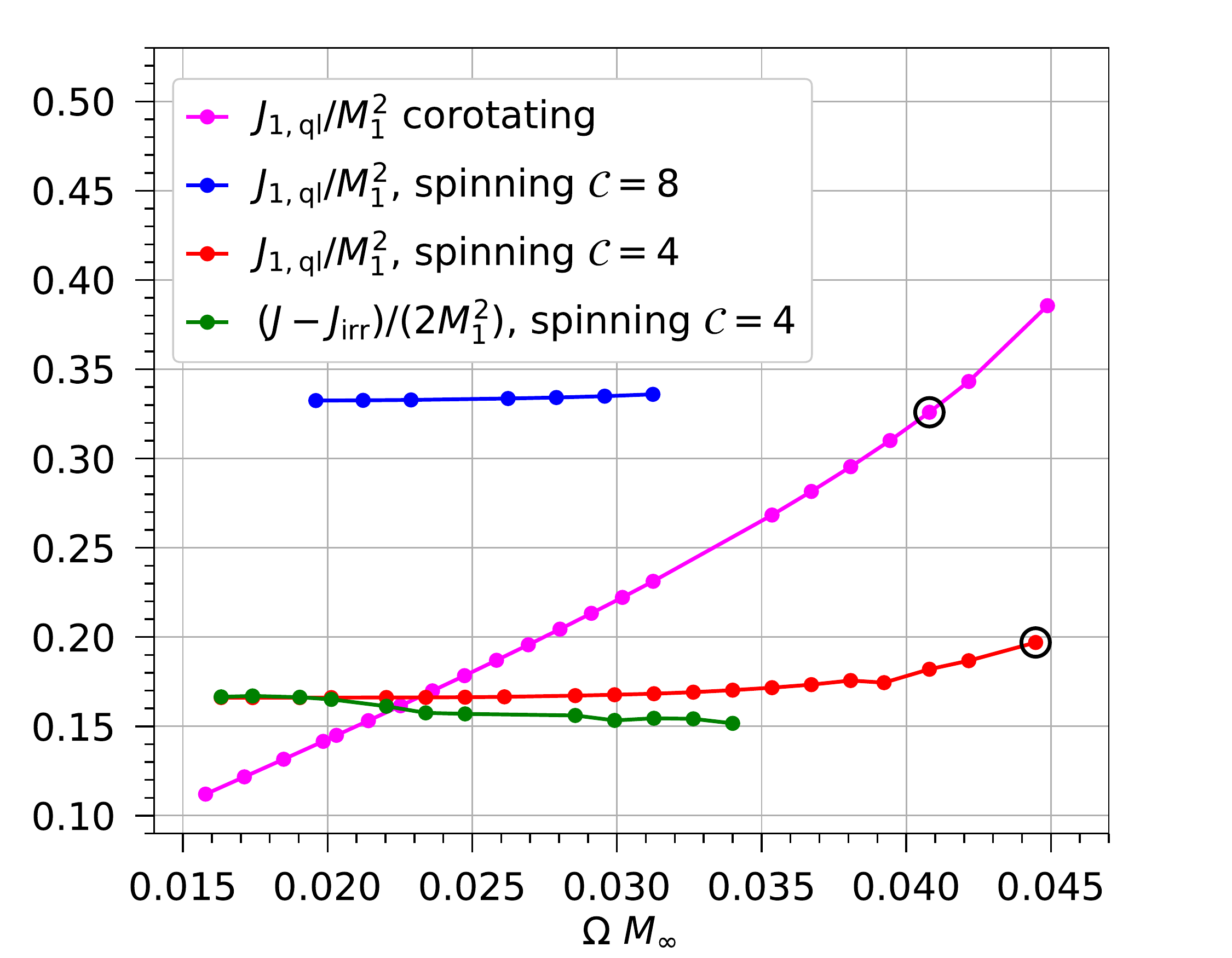}
\caption{Spin measures for an individual star in a binary setting. $M_1$
corresponds to the ADM mass of a single star at infinity, and $J_{1,\rm ql}$
corresponds to its quasilocal spin. Except for the green curve, all others show
the quasilocal spin of a single star along a sequence. The green curve estimates
the spin by comparison with an irrotational sequence at the same orbital angular
velocity points. Points marked with a larger black circle denote the
approximate ISCO.  }  
\label{fig:ALF2Jql}
\end{center}
\end{figure}

In Fig. \ref{fig:ALF2binJM2}, we plot the total angular momentum of the system
for a sequence of constant circulation $\mathcal{C}=4$, together with the
familiar irrotational and corotating sequences.  Also, the corresponding PN
curves are plotted. The qualitative feature of a constant circulation curve is
that it runs parallel to the irrotational curve at a higher angular
momentum level for aligned spin binaries. This is not surprising since an
irrotational curve has constant circulation $\mathcal{C}=0$. Higher spinning
binaries have curves shifted upward, and antialigned spinning binaries have
curves parallel and below the level of the irrotational one. Another feature is
that at large separations the constant circulation curve does not converge to
the PN curves, which is also expected since these binaries have spin angular
momentum independent of the orbital angular momentum. That is also the reason
why they intersect the corotating sequence curve which has small spin angular
momentum at infinity and becomes larger as one moves toward smaller distances.
Given the fact that a dynamical evolution conserves the rest mass, entropy, and
circulation, a physical spinning sequence representing a merging binary is going
to be like the red or blue one in Fig. \ref{fig:ALF2binJM2}.  Points marked with
a larger black circle denote the approximate innermost stable circular orbit
(ISCO). Locating the ISCO is not essential for this work therefore its location
as denoted in Figs. \ref{fig:ALF2binJM2},\ref{fig:ALF2Jql} can be further
refined.

In Fig. \ref{fig:ALF2Jql}, different spin measures are plotted along constant
circulation sequences as well as a corotating one. $M_1=1.36$ corresponds to the
ADM mass of a single star at infinity, and $J_{1,\rm ql}$ corresponds to its
quasilocal spin as calculated from Eq. (\ref{eq:qlspin}) but with the rotational
vector $\GP^i_{\rm s}$ (which generates rotations around the star's center)
instead of $\GP^i$.  $J$ is the total angular momentum of the binary system, and
$J_{\rm irr}$ is the total angular momentum of the irrotational binary at the
same angular velocity. From the corotating (purple) sequence, one can see that
the dimensionless spin $J_{1,\rm ql}/M_1^2$ grows linearly as the separation
decreases.  \re{Also even at very close separation (ISCO) this dimensionless
spin is relatively small $< 0.35$}.  This linear growth of the quasilocal spin
is consistent with Figs. \ref{fig:circ_jm2} (and \ref{fig:cocirc}), which also
shows that behavior for small $J/M^2$ in single rotating stars. Sequences of
constant circulation $\mathcal{C}=4,8$ are also plotted in Fig.
\ref{fig:ALF2Jql}. The curves (blue and red) show that within the accuracy of
our computation the dimensionless quasilocal spin (or equivalently the
quasilocal angular momentum) is also conserved along these sequences when the
binaries are widely separated. As one moves towards the ISCO we observe a $\sim
10-15\%$ increase which is consistent with the increase found in evolutions
\cite{Dietrich:2016lyp}.  This behavior is also consistent with Fig.
\ref{fig:ALF2star}, which shows that for slowly rotating single stars sequences
of constant circulation are parallel to sequences of constant $J/M^2$. Another
measure of spin typically quoted in the literature is the difference between the
angular momentum at infinity of the irrotational solution from the corresponding
spinning solution.  In Fig. \ref{fig:ALF2Jql}, we plot this spin measure of the
$\mathcal{C}=4$ sequence by comparing it with the corresponding irrotational
sequence (green curve). The plot shows that, although at larger separations the
two diagnostics agree with each other, as one moves to closer separations they
start to diverge. This is to be expected since the $J-J_{\rm irr}$ angular
momentum contains negative terms (1.5 PN) related to the spin orbit coupling
\cite{2014LRR....17....2B}.

\begin{figure}
\begin{center}
\includegraphics[width=0.99\columnwidth]{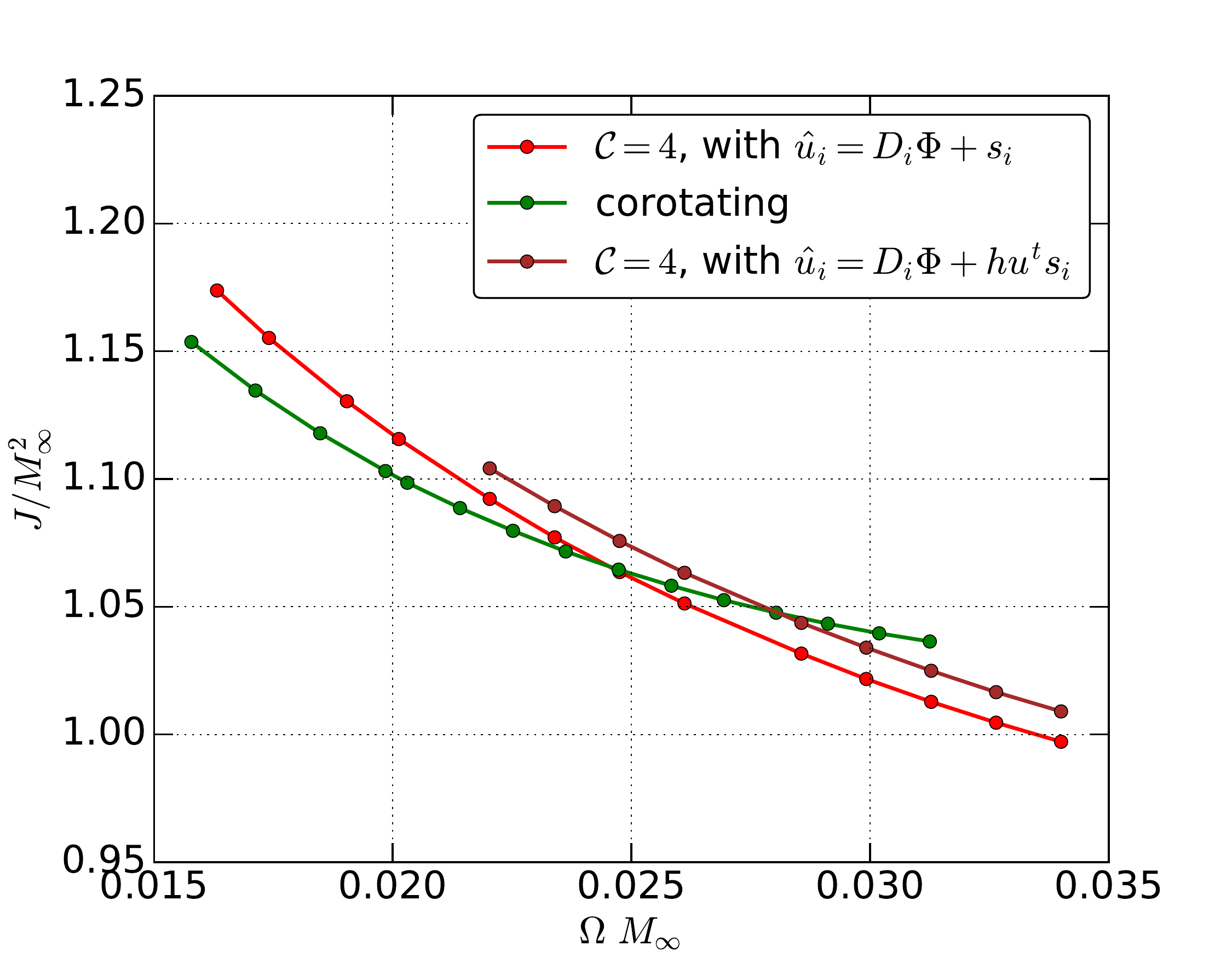}
\caption{Angular momentum curve for a binary system with constant circulation
$\mathcal{C}=4$ and rest mass $M_0=1.5$, using decomposition  (\ref{eq:uihnew}),
along with the same sequence as presented in Fig. \ref{fig:ALF2binJM2}, which
uses the original decomposition Eq. (\ref{eq:uih_dec}). Also shown is the
corotating sequence.}  
\label{fig:ALF2binJM2new}
\end{center}
\end{figure}

\section{Modified spin formulation}

Motivated by the circulation expression for single stars and corotating binaries
(\ref{eq:circstar}),
we investigate a modification for the decomposition  (\ref{eq:uih_dec}) 
proposed by Tichy \cite{Tichy:2011gw}; i.e. we take 
\be
\hat{u}_i := \GG_i^\GA hu_\GA = D_i\Phi + h u^t s_i    \label{eq:uihnew}
\ee
but otherwise adopt the same assumptions. In doing so, the circulation of a
spinning star in a binary will be
\be
\mathcal{C} = R_0 \oint_c h u^t \GC^4 \GD_{ij}\hat{\Omega}_{\rm s} \hat{\GP}^i_{\rm s} d\hat{x}^i \,  ,
\label{eq:newcirc}
\ee
which apart from the coordinate terms (due to shift $\GB^i$) closely matches Eq.
(\ref{eq:circstar}) of the circulation of a single rotating star. Now, the
velocity with respect to the corotating frame becomes 
\be
 V^i=\frac{D^i\Phi}{hu^t}-(\GO^i-s^i)   \, , \label{eq:Vnew}
\ee
which can be thought as the same with the irrotational case and a replace 
\be
\GO^i \quad\longleftrightarrow\quad  \GO^i - s^i  \, , \label{eq:rep}
\ee
where again here $\GO^i=\GB^i+\Omega\GP^i$ is the corotating shift. The Euler
first integral now becomes
\be
h^2 + D_i \Phi D^i\Phi = \GL hu^t   \label{eq:Eulernew}
\ee
where $\GL:=C+(\GO^i - s^i)D_i\Phi$.  It turns out now that the equations are
simplified and the relative quantities can be computed through a linear equation
in $hu^t$,
\be
hu^t = \frac{\GL+2s^i D_i\Phi}{\GA^2 -s_i s^i} \ . \label{eq:hutnew}
\ee
The denominator in the expression above is larger than zero, since even for very
compact stars $\GA^2>0.1$, which is approximately 1 order of magnitude larger
than the square of the spin magnitude.  Once $hu^t$ is computed from Eq.
(\ref{eq:hutnew}), the enthalpy is calculated from Eq. (\ref{eq:Eulernew}).

The velocity potential is determined from the conservation of rest mass, 
\begin{eqnarray}
\nabla^2\Phi & = & -\frac{2}{\GC}\pd^i\GC\pd_i\Phi +\GC^4\pd_i[hu^t(\GO^i-s^i)] \nonumber\\
             & + & 6hu^t\GC^3(\GO^i-s^i)\pd_i\GC \nonumber\\
             & - & \pd_i\ln\left(\frac{\GA\GP}{h}\right)[\pd^i\Phi-\GC^4hu^t(\GO^i-s^i)]
\end{eqnarray}
with boundary condition
\be
\{[-\pd^i\Phi+\GC^4 h u^t (\GO^i-s^i)]\pd_i\GR\}_{\textrm{surface}} = 0 .
\ee

In Fig. \ref{fig:ALF2binJM2new}, we plot a sequence of constant rest mass
$M_0=1.5$ and constant circulation $\mathcal{C}=4$ using decomposition
(\ref{eq:uihnew})  along with the same sequence using the original decomposition
(\ref{eq:uih_dec}) that we plotted in Fig. \ref{fig:ALF2binJM2}.  We also show
the corotating sequence for comparison.  It is evident that the way one
decomposes the velocity $\hat{u}_i$ introduces an arbitrariness in the
circulation, which in the present case results in a higher angular momentum for
the system. This is not difficult to explain since the parabolic functional form
of the $hu^t$ factor in Eq. (\ref{eq:uihnew}) results in a differentially
rotating BNS, which increases the angular momentum of the system. On the other
hand, this differential rotation, which naturally results from Eq.
(\ref{eq:uihnew}), can be canceled or modified by an appropriate choice of the
input vector $s^i$, which  must have a varying parameter $\Omega_s$.  Since
spinning BNSs are expected to have a rotation law which is close to rigid
rotation, decomposition (\ref{eq:uih_dec}) is closer to astrophysical
expectations over (\ref{eq:uihnew}).  The latter can still produce almost
uniformly rotating objects, but the spin input vector $s_i$ is nontrivial.

\section{Discussion}

Dynamical evolution of isentropic fluids conserves the baryon mass, entropy, and
vorticity. Therefore, along with the rest mass, one can use the circulation of a
neutron star to compute realistic sequences of binary neutron stars and measure
their individual spin. In this paper, we extended our \cocal{} code to compute
such equilibria and used it to make a critical assessment of various spin
measures for BNS, as well as a connection with the spin of single rotating
stars. 

By computing sequences of constant angular momentum $J$, angular velocity,
circulation, and dimensionless spin $J/M^2$ for single axisymmetric stars, we
showed that in general all such family curves are distinct.  For small spins,
though, curves of constant circulation ``run parallel'' to those of constant
$J/M^2$; therefore, conservation of circulation implies conservation of $J/M^2$
and vice versa. Using the approximation of conformal flatness and maximal
slicing (which is typically used for BNS calculations), the angular momentum $J$
equals the quasilocal spin $J_{\rm ql}$, which is widely used to measure the
angular momentum of a compact body in a binary scenario.  For BNSs, neighboring
equilibria satisfy the first law of thermodynamics by Friedman \textit{et al.}
and by computing sequences of constant rest mass and circulation, we show that
the dimensionless spin is also approximately conserved at least for low spin
binaries. 

Motivated by the expression of circulation in single rotating stars, we explored
an alternative decomposition for the 4-velocity than the one originally proposed
by Tichy, which naturally led to differentially rotating binary systems, and
discussed a potential ambiguity that results from any such decomposition.

\textit{Acknowledgments.}\textemdash 
This work was supported by NSF Grants No.  PHY-1602536 and No. PHY-1662211 and
NASA Grant No.  80NSSC17K0070 to the University of Illinois at Urbana-Champaign
as well as by a  JSPS Grant-in-Aid for Scientific Research (C), Grants No.
15K05085 and No.  18K03624, to the University of Ryukyus.

\appendix*

\section{Spin parameter $\Omega_{\rm s}$ along a constant circulation
and rest-mass sequence}

\begin{figure}
\begin{center}
\includegraphics[width=0.99\columnwidth]{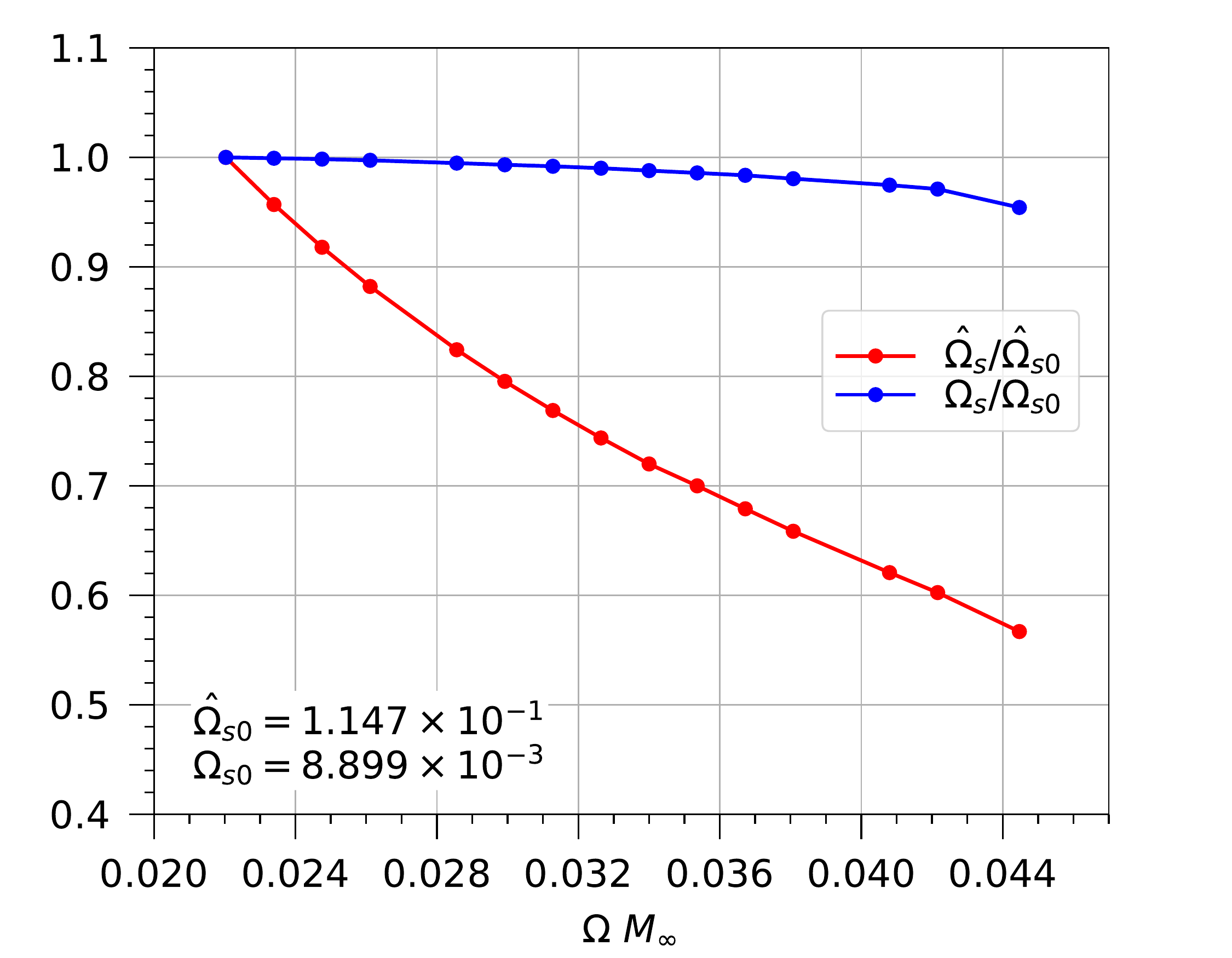}
\caption{Spin parameter $\Omega_s$ and the normalized parameter $\hat{\Omega}_s$
along the constant circulation $\mathcal{C}=4$ and the constant rest mass
$M_0=1.5$ sequence of Fig. \ref{fig:ALF2binJM2}. }  
\label{fig:ALF2omes}
\end{center}
\end{figure}

In Fig. \ref{fig:ALF2omes}, we plot the spin parameter, both the original
$\Omega_{\rm s}$ and the normalized one $\hat{\Omega}_s$, for the
$\mathcal{C}=4$ sequence. To a high degree, a constant circulation sequence
corresponds to a constant spinning parameter $\Omega_s$ for widely separated
binaries [see Eq. (\ref{eq:spin})], but the normalized parameter
$\hat{\Omega}_s$, which is used in our code, varies considerably along the
sequence.  \re{Along a constant circulation sequence the maximum variation of
$\Omega_s$ happens at the ISCO and is $\sim 4\%$.} Having said that, we must
keep in mind that Fig. \ref{fig:ALF2omes} corresponds to $\mathcal{C}=4$ or
according to Fig. \ref{fig:ALF2Jql} quasilocal spin of $\sim 0.17$. For high
enough spins ($>0.5$), this behavior may not be true. Also, if for the spin
vector $s^i$, Eq. (\ref{eq:spin}), one uses a more complicated expression (for
example with multiple parameters), the behavior can change analogously.  \re{For
the new sequence plotted in Fig. \ref{fig:ALF2binJM2new} using decomposition Eq.
(\ref{eq:uihnew}), the variation of $\Omega_s$ is twice of that of Fig.
\ref{fig:ALF2omes} using Eq. (\ref{eq:uih_dec}). } In other words, the
decomposition (\ref{eq:uih_dec}) introduces an arbitrariness to $\hat{u}_i$
through the input spin vector $s^i$, which is necessary for computing the
circulation. In a realistic scenario, any given spinning BNS has a particular
$\hat{u}_i$, which is the result of hydrostatic equilibrium and its evolutionary
history, and this determines its circulation.  Targeting the circulation alone
does not uniquely specify the velocity profile in the configuration.  Hence, we
can construct two sequences with the same circulation, one with a constant and
the other with differential angular velocity, as we have seen in the last
section above.

\vspace{4cm}




\bibliographystyle{apsrev4-1}
\bibliography{references}

\begin{thebibliography}{46}%
\makeatletter
\providecommand \@ifxundefined [1]{%
 \@ifx{#1\undefined}
}%
\providecommand \@ifnum [1]{%
 \ifnum #1\expandafter \@firstoftwo
 \else \expandafter \@secondoftwo
 \fi
}%
\providecommand \@ifx [1]{%
 \ifx #1\expandafter \@firstoftwo
 \else \expandafter \@secondoftwo
 \fi
}%
\providecommand \natexlab [1]{#1}%
\providecommand \enquote  [1]{``#1''}%
\providecommand \bibnamefont  [1]{#1}%
\providecommand \bibfnamefont [1]{#1}%
\providecommand \citenamefont [1]{#1}%
\providecommand \href@noop [0]{\@secondoftwo}%
\providecommand \href [0]{\begingroup \@sanitize@url \@href}%
\providecommand \@href[1]{\@@startlink{#1}\@@href}%
\providecommand \@@href[1]{\endgroup#1\@@endlink}%
\providecommand \@sanitize@url [0]{\catcode `\\12\catcode `\$12\catcode
  `\&12\catcode `\#12\catcode `\^12\catcode `\_12\catcode `\%12\relax}%
\providecommand \@@startlink[1]{}%
\providecommand \@@endlink[0]{}%
\providecommand \url  [0]{\begingroup\@sanitize@url \@url }%
\providecommand \@url [1]{\endgroup\@href {#1}{\urlprefix }}%
\providecommand \urlprefix  [0]{URL }%
\providecommand \Eprint [0]{\href }%
\providecommand \doibase [0]{http://dx.doi.org/}%
\providecommand \selectlanguage [0]{\@gobble}%
\providecommand \bibinfo  [0]{\@secondoftwo}%
\providecommand \bibfield  [0]{\@secondoftwo}%
\providecommand \translation [1]{[#1]}%
\providecommand \BibitemOpen [0]{}%
\providecommand \bibitemStop [0]{}%
\providecommand \bibitemNoStop [0]{.\EOS\space}%
\providecommand \EOS [0]{\spacefactor3000\relax}%
\providecommand \BibitemShut  [1]{\csname bibitem#1\endcsname}%
\let\auto@bib@innerbib\@empty
\bibitem [{\citenamefont {{Abbott}}\ \emph {et~al.}(2017)\citenamefont
  {{Abbott}} \emph {et~al.}}]{2017PhRvL.119p1101A}%
  \BibitemOpen
  \bibfield  {author} {\bibinfo {author} {\bibfnamefont {B.~P.}\ \bibnamefont
  {{Abbott}}} \emph {et~al.},\ }\href {\doibase 10.1103/PhysRevLett.119.161101}
  {\bibfield  {journal} {\bibinfo  {journal} {Phys. Rev. Lett.}\ }\textbf
  {\bibinfo {volume} {119}},\ \bibinfo {eid} {161101} (\bibinfo {year}
  {2017})},\ \Eprint {http://arxiv.org/abs/1710.05832} {arXiv:1710.05832
  [gr-qc]} \BibitemShut {NoStop}%
\bibitem [{\citenamefont {Goldstein}\ \emph {et~al.}(2017)\citenamefont
  {Goldstein} \emph {et~al.}}]{2041-8205-848-2-L14}%
  \BibitemOpen
  \bibfield  {author} {\bibinfo {author} {\bibfnamefont {A.}~\bibnamefont
  {Goldstein}} \emph {et~al.},\ }\href
  {http://stacks.iop.org/2041-8205/848/i=2/a=L14} {\bibfield  {journal}
  {\bibinfo  {journal} {The Astrophys. J. Lett.}\ }\textbf {\bibinfo {volume}
  {848}},\ \bibinfo {pages} {L14} (\bibinfo {year} {2017})}\BibitemShut
  {NoStop}%
\bibitem [{\citenamefont {Savchenko}\ \emph {et~al.}(2017)\citenamefont
  {Savchenko} \emph {et~al.}}]{Savchenko:2017ffs}%
  \BibitemOpen
  \bibfield  {author} {\bibinfo {author} {\bibfnamefont {V.}~\bibnamefont
  {Savchenko}} \emph {et~al.},\ }\href {\doibase 10.3847/2041-8213/aa8f94}
  {\bibfield  {journal} {\bibinfo  {journal} {Astrophys. J.}\ }\textbf
  {\bibinfo {volume} {848}},\ \bibinfo {pages} {L15} (\bibinfo {year}
  {2017})},\ \Eprint {http://arxiv.org/abs/1710.05449} {arXiv:1710.05449
  [astro-ph.HE]} \BibitemShut {NoStop}%
\bibitem [{\citenamefont {{Hessels}}\ \emph {et~al.}(2006)\citenamefont
  {{Hessels}}, \citenamefont {{Ransom}}, \citenamefont {{Stairs}},
  \citenamefont {{Freire}}, \citenamefont {{Kaspi}},\ and\ \citenamefont
  {{Camilo}}}]{2006Sci...311.1901H}%
  \BibitemOpen
  \bibfield  {author} {\bibinfo {author} {\bibfnamefont {J.~W.~T.}\
  \bibnamefont {{Hessels}}}, \bibinfo {author} {\bibfnamefont {S.~M.}\
  \bibnamefont {{Ransom}}}, \bibinfo {author} {\bibfnamefont {I.~H.}\
  \bibnamefont {{Stairs}}}, \bibinfo {author} {\bibfnamefont {P.~C.~C.}\
  \bibnamefont {{Freire}}}, \bibinfo {author} {\bibfnamefont {V.~M.}\
  \bibnamefont {{Kaspi}}}, \ and\ \bibinfo {author} {\bibfnamefont
  {F.}~\bibnamefont {{Camilo}}},\ }\href {\doibase 10.1126/science.1123430}
  {\bibfield  {journal} {\bibinfo  {journal} {Science}\ }\textbf {\bibinfo
  {volume} {311}},\ \bibinfo {pages} {1901} (\bibinfo {year} {2006})},\ \Eprint
  {http://arxiv.org/abs/astro-ph/0601337} {astro-ph/0601337} \BibitemShut
  {NoStop}%
\bibitem [{\citenamefont {Tauris}\ \emph {et~al.}(2017)\citenamefont {Tauris}
  \emph {et~al.}}]{Tauris:2017omb}%
  \BibitemOpen
  \bibfield  {author} {\bibinfo {author} {\bibfnamefont {T.~M.}\ \bibnamefont
  {Tauris}} \emph {et~al.},\ }\href {\doibase 10.3847/1538-4357/aa7e89}
  {\bibfield  {journal} {\bibinfo  {journal} {Astrophys. J.}\ }\textbf
  {\bibinfo {volume} {846}},\ \bibinfo {pages} {170} (\bibinfo {year}
  {2017})},\ \Eprint {http://arxiv.org/abs/1706.09438} {arXiv:1706.09438
  [astro-ph.HE]} \BibitemShut {NoStop}%
\bibitem [{\citenamefont {Zhu}\ \emph {et~al.}(2017)\citenamefont {Zhu},
  \citenamefont {Thrane}, \citenamefont {Oslowski}, \citenamefont {Levin},\
  and\ \citenamefont {Lasky}}]{Zhu:2017znf}%
  \BibitemOpen
  \bibfield  {author} {\bibinfo {author} {\bibfnamefont {X.}~\bibnamefont
  {Zhu}}, \bibinfo {author} {\bibfnamefont {E.}~\bibnamefont {Thrane}},
  \bibinfo {author} {\bibfnamefont {S.}~\bibnamefont {Oslowski}}, \bibinfo
  {author} {\bibfnamefont {Y.}~\bibnamefont {Levin}}, \ and\ \bibinfo {author}
  {\bibfnamefont {P.~D.}\ \bibnamefont {Lasky}},\ }\href@noop {} {\  (\bibinfo
  {year} {2017})},\ \Eprint {http://arxiv.org/abs/1711.09226} {arXiv:1711.09226
  [astro-ph.HE]} \BibitemShut {NoStop}%
\bibitem [{\citenamefont {Stovall}\ \emph {et~al.}(2018)\citenamefont {Stovall}
  \emph {et~al.}}]{Stovall:2018ouw}%
  \BibitemOpen
  \bibfield  {author} {\bibinfo {author} {\bibfnamefont {K.}~\bibnamefont
  {Stovall}} \emph {et~al.},\ }\href {\doibase 10.3847/2041-8213/aaad06}
  {\bibfield  {journal} {\bibinfo  {journal} {Astrophys. J.}\ }\textbf
  {\bibinfo {volume} {854}},\ \bibinfo {pages} {L22} (\bibinfo {year}
  {2018})},\ \Eprint {http://arxiv.org/abs/1802.01707} {arXiv:1802.01707
  [astro-ph.HE]} \BibitemShut {NoStop}%
\bibitem [{\citenamefont {Cameron}\ \emph {et~al.}(2018)\citenamefont {Cameron}
  \emph {et~al.}}]{Cameron:2017ody}%
  \BibitemOpen
  \bibfield  {author} {\bibinfo {author} {\bibfnamefont {A.~D.}\ \bibnamefont
  {Cameron}} \emph {et~al.},\ }\href {\doibase 10.1093/mnrasl/sly003}
  {\bibfield  {journal} {\bibinfo  {journal} {Mon. Not. R. Astron. Soc.}\
  }\textbf {\bibinfo {volume} {475}},\ \bibinfo {pages} {L57} (\bibinfo {year}
  {2018})},\ \Eprint {http://arxiv.org/abs/1711.07697} {arXiv:1711.07697
  [astro-ph.HE]} \BibitemShut {NoStop}%
\bibitem [{\citenamefont {Kramer}\ \emph {et~al.}(2006)\citenamefont {Kramer}
  \emph {et~al.}}]{Kramer:2006nb}%
  \BibitemOpen
  \bibfield  {author} {\bibinfo {author} {\bibfnamefont {M.}~\bibnamefont
  {Kramer}} \emph {et~al.},\ }\href {\doibase 10.1126/science.1132305}
  {\bibfield  {journal} {\bibinfo  {journal} {Science}\ }\textbf {\bibinfo
  {volume} {314}},\ \bibinfo {pages} {97} (\bibinfo {year} {2006})},\ \Eprint
  {http://arxiv.org/abs/astro-ph/0609417} {arXiv:astro-ph/0609417 [astro-ph]}
  \BibitemShut {NoStop}%
\bibitem [{\citenamefont {{Baumgarte}}\ \emph {et~al.}(1997)\citenamefont
  {{Baumgarte}}, \citenamefont {{Cook}}, \citenamefont {{Scheel}},
  \citenamefont {{Shapiro}},\ and\ \citenamefont
  {{Teukolsky}}}]{1997PhRvL..79.1182B}%
  \BibitemOpen
  \bibfield  {author} {\bibinfo {author} {\bibfnamefont {T.~W.}\ \bibnamefont
  {{Baumgarte}}}, \bibinfo {author} {\bibfnamefont {G.~B.}\ \bibnamefont
  {{Cook}}}, \bibinfo {author} {\bibfnamefont {M.~A.}\ \bibnamefont
  {{Scheel}}}, \bibinfo {author} {\bibfnamefont {S.~L.}\ \bibnamefont
  {{Shapiro}}}, \ and\ \bibinfo {author} {\bibfnamefont {S.~A.}\ \bibnamefont
  {{Teukolsky}}},\ }\href {\doibase 10.1103/PhysRevLett.79.1182} {\bibfield
  {journal} {\bibinfo  {journal} {\prl}\ }\textbf {\bibinfo {volume} {79}},\
  \bibinfo {pages} {1182} (\bibinfo {year} {1997})},\ \Eprint
  {http://arxiv.org/abs/gr-qc/9704024} {arXiv:gr-qc/9704024} \BibitemShut
  {NoStop}%
\bibitem [{\citenamefont {{Baumgarte}}\ \emph {et~al.}(1998)\citenamefont
  {{Baumgarte}}, \citenamefont {{Cook}}, \citenamefont {{Scheel}},
  \citenamefont {{Shapiro}},\ and\ \citenamefont
  {{Teukolsky}}}]{1998PhRvD..57.7299B}%
  \BibitemOpen
  \bibfield  {author} {\bibinfo {author} {\bibfnamefont {T.~W.}\ \bibnamefont
  {{Baumgarte}}}, \bibinfo {author} {\bibfnamefont {G.~B.}\ \bibnamefont
  {{Cook}}}, \bibinfo {author} {\bibfnamefont {M.~A.}\ \bibnamefont
  {{Scheel}}}, \bibinfo {author} {\bibfnamefont {S.~L.}\ \bibnamefont
  {{Shapiro}}}, \ and\ \bibinfo {author} {\bibfnamefont {S.~A.}\ \bibnamefont
  {{Teukolsky}}},\ }\href {\doibase 10.1103/PhysRevD.57.7299} {\bibfield
  {journal} {\bibinfo  {journal} {\prd}\ }\textbf {\bibinfo {volume} {57}},\
  \bibinfo {pages} {7299} (\bibinfo {year} {1998})},\ \Eprint
  {http://arxiv.org/abs/gr-qc/9709026} {arXiv:gr-qc/9709026} \BibitemShut
  {NoStop}%
\bibitem [{\citenamefont {{Marronetti}}\ \emph {et~al.}(1998)\citenamefont
  {{Marronetti}}, \citenamefont {{Mathews}},\ and\ \citenamefont
  {{Wilson}}}]{1998PhRvD..58j7503M}%
  \BibitemOpen
  \bibfield  {author} {\bibinfo {author} {\bibfnamefont {P.}~\bibnamefont
  {{Marronetti}}}, \bibinfo {author} {\bibfnamefont {G.~J.}\ \bibnamefont
  {{Mathews}}}, \ and\ \bibinfo {author} {\bibfnamefont {J.~R.}\ \bibnamefont
  {{Wilson}}},\ }\href {\doibase 10.1103/PhysRevD.58.107503} {\bibfield
  {journal} {\bibinfo  {journal} {\prd}\ }\textbf {\bibinfo {volume} {58}},\
  \bibinfo {eid} {107503} (\bibinfo {year} {1998})},\ \Eprint
  {http://arxiv.org/abs/gr-qc/9803093} {arXiv:gr-qc/9803093} \BibitemShut
  {NoStop}%
\bibitem [{\citenamefont {{Bildsten}}\ and\ \citenamefont
  {{Cutler}}(1992)}]{1992ApJ...400..175B}%
  \BibitemOpen
  \bibfield  {author} {\bibinfo {author} {\bibfnamefont {L.}~\bibnamefont
  {{Bildsten}}}\ and\ \bibinfo {author} {\bibfnamefont {C.}~\bibnamefont
  {{Cutler}}},\ }\href {\doibase 10.1086/171983} {\bibfield  {journal}
  {\bibinfo  {journal} {\apj}\ }\textbf {\bibinfo {volume} {400}},\ \bibinfo
  {pages} {175} (\bibinfo {year} {1992})}\BibitemShut {NoStop}%
\bibitem [{\citenamefont {{Kochanek}}(1992)}]{1992ApJ...398..234K}%
  \BibitemOpen
  \bibfield  {author} {\bibinfo {author} {\bibfnamefont {C.~S.}\ \bibnamefont
  {{Kochanek}}},\ }\href {\doibase 10.1086/171851} {\bibfield  {journal}
  {\bibinfo  {journal} {\apj}\ }\textbf {\bibinfo {volume} {398}},\ \bibinfo
  {pages} {234} (\bibinfo {year} {1992})}\BibitemShut {NoStop}%
\bibitem [{\citenamefont {{Bonazzola}}\ \emph {et~al.}(1999)\citenamefont
  {{Bonazzola}}, \citenamefont {{Gourgoulhon}},\ and\ \citenamefont
  {{Marck}}}]{1999PhRvL..82..892B}%
  \BibitemOpen
  \bibfield  {author} {\bibinfo {author} {\bibfnamefont {S.}~\bibnamefont
  {{Bonazzola}}}, \bibinfo {author} {\bibfnamefont {E.}~\bibnamefont
  {{Gourgoulhon}}}, \ and\ \bibinfo {author} {\bibfnamefont {J.-A.}\
  \bibnamefont {{Marck}}},\ }\href {\doibase 10.1103/PhysRevLett.82.892}
  {\bibfield  {journal} {\bibinfo  {journal} {\prl}\ }\textbf {\bibinfo
  {volume} {82}},\ \bibinfo {pages} {892} (\bibinfo {year} {1999})},\ \Eprint
  {http://arxiv.org/abs/gr-qc/9810072} {arXiv:gr-qc/9810072} \BibitemShut
  {NoStop}%
\bibitem [{\citenamefont {{Gourgoulhon}}\ \emph {et~al.}(2001)\citenamefont
  {{Gourgoulhon}}, \citenamefont {{Grandcl{\'e}ment}}, \citenamefont
  {{Taniguchi}}, \citenamefont {{Marck}},\ and\ \citenamefont
  {{Bonazzola}}}]{2001PhRvD..63f4029G}%
  \BibitemOpen
  \bibfield  {author} {\bibinfo {author} {\bibfnamefont {E.}~\bibnamefont
  {{Gourgoulhon}}}, \bibinfo {author} {\bibfnamefont {P.}~\bibnamefont
  {{Grandcl{\'e}ment}}}, \bibinfo {author} {\bibfnamefont {K.}~\bibnamefont
  {{Taniguchi}}}, \bibinfo {author} {\bibfnamefont {J.-A.}\ \bibnamefont
  {{Marck}}}, \ and\ \bibinfo {author} {\bibfnamefont {S.}~\bibnamefont
  {{Bonazzola}}},\ }\href {\doibase 10.1103/PhysRevD.63.064029} {\bibfield
  {journal} {\bibinfo  {journal} {\prd}\ }\textbf {\bibinfo {volume} {63}},\
  \bibinfo {eid} {064029} (\bibinfo {year} {2001})},\ \Eprint
  {http://arxiv.org/abs/gr-qc/0007028} {arXiv:gr-qc/0007028} \BibitemShut
  {NoStop}%
\bibitem [{\citenamefont {{Marronetti}}\ \emph {et~al.}(2000)\citenamefont
  {{Marronetti}}, \citenamefont {{Mathews}},\ and\ \citenamefont
  {{Wilson}}}]{2000NuPhS..80C0714M}%
  \BibitemOpen
  \bibfield  {author} {\bibinfo {author} {\bibfnamefont {P.}~\bibnamefont
  {{Marronetti}}}, \bibinfo {author} {\bibfnamefont {G.~J.}\ \bibnamefont
  {{Mathews}}}, \ and\ \bibinfo {author} {\bibfnamefont {J.~R.}\ \bibnamefont
  {{Wilson}}},\ }\href@noop {} {\bibfield  {journal} {\bibinfo  {journal}
  {Nucl. Phys. B}\ }\textbf {\bibinfo {volume} {80}},\ \bibinfo {pages} {07/14}
  (\bibinfo {year} {2000})},\ \Eprint {http://arxiv.org/abs/gr-qc/9903105}
  {arXiv:gr-qc/9903105} \BibitemShut {NoStop}%
\bibitem [{\citenamefont {{Marronetti}}\ \emph {et~al.}(1999)\citenamefont
  {{Marronetti}}, \citenamefont {{Mathews}},\ and\ \citenamefont
  {{Wilson}}}]{1999PhRvD..60h7301M}%
  \BibitemOpen
  \bibfield  {author} {\bibinfo {author} {\bibfnamefont {P.}~\bibnamefont
  {{Marronetti}}}, \bibinfo {author} {\bibfnamefont {G.~J.}\ \bibnamefont
  {{Mathews}}}, \ and\ \bibinfo {author} {\bibfnamefont {J.~R.}\ \bibnamefont
  {{Wilson}}},\ }\href {\doibase 10.1103/PhysRevD.60.087301} {\bibfield
  {journal} {\bibinfo  {journal} {\prd}\ }\textbf {\bibinfo {volume} {60}},\
  \bibinfo {eid} {087301} (\bibinfo {year} {1999})},\ \Eprint
  {http://arxiv.org/abs/gr-qc/9906088} {arXiv:gr-qc/9906088} \BibitemShut
  {NoStop}%
\bibitem [{\citenamefont {{Ury{\={u}}}}\ and\ \citenamefont
  {{Eriguchi}}(2000)}]{2000PhRvD..61l4023U}%
  \BibitemOpen
  \bibfield  {author} {\bibinfo {author} {\bibfnamefont {K.}~\bibnamefont
  {{Ury{\={u}}}}}\ and\ \bibinfo {author} {\bibfnamefont {Y.}~\bibnamefont
  {{Eriguchi}}},\ }\href {\doibase 10.1103/PhysRevD.61.124023} {\bibfield
  {journal} {\bibinfo  {journal} {\prd}\ }\textbf {\bibinfo {volume} {61}},\
  \bibinfo {eid} {124023} (\bibinfo {year} {2000})},\ \Eprint
  {http://arxiv.org/abs/gr-qc/9908059} {arXiv:gr-qc/9908059} \BibitemShut
  {NoStop}%
\bibitem [{\citenamefont {{Ury{\={u}}}}\ \emph {et~al.}(2000)\citenamefont
  {{Ury{\={u}}}}, \citenamefont {{Shibata}},\ and\ \citenamefont
  {{Eriguchi}}}]{2000PhRvD..62j4015U}%
  \BibitemOpen
  \bibfield  {author} {\bibinfo {author} {\bibfnamefont {K.}~\bibnamefont
  {{Ury{\={u}}}}}, \bibinfo {author} {\bibfnamefont {M.}~\bibnamefont
  {{Shibata}}}, \ and\ \bibinfo {author} {\bibfnamefont {Y.}~\bibnamefont
  {{Eriguchi}}},\ }\href {\doibase 10.1103/PhysRevD.62.104015} {\bibfield
  {journal} {\bibinfo  {journal} {\prd}\ }\textbf {\bibinfo {volume} {62}},\
  \bibinfo {eid} {104015} (\bibinfo {year} {2000})},\ \Eprint
  {http://arxiv.org/abs/gr-qc/0007042} {arXiv:gr-qc/0007042} \BibitemShut
  {NoStop}%
\bibitem [{\citenamefont {Marronetti}\ and\ \citenamefont
  {Shapiro}(2003)}]{Marronetti:2003gk}%
  \BibitemOpen
  \bibfield  {author} {\bibinfo {author} {\bibfnamefont {P.}~\bibnamefont
  {Marronetti}}\ and\ \bibinfo {author} {\bibfnamefont {S.~L.}\ \bibnamefont
  {Shapiro}},\ }\href {\doibase 10.1103/PhysRevD.68.104024} {\bibfield
  {journal} {\bibinfo  {journal} {\prd}\ }\textbf {\bibinfo {volume} {68}},\
  \bibinfo {pages} {104024} (\bibinfo {year} {2003})},\ \Eprint
  {http://arxiv.org/abs/gr-qc/0306075} {arXiv:gr-qc/0306075 [gr-qc]}
  \BibitemShut {NoStop}%
\bibitem [{\citenamefont {{Baumgarte}}\ and\ \citenamefont
  {{Shapiro}}(2009{\natexlab{a}})}]{2009PhRvD..80f4009B}%
  \BibitemOpen
  \bibfield  {author} {\bibinfo {author} {\bibfnamefont {T.~W.}\ \bibnamefont
  {{Baumgarte}}}\ and\ \bibinfo {author} {\bibfnamefont {S.~L.}\ \bibnamefont
  {{Shapiro}}},\ }\href {\doibase 10.1103/PhysRevD.80.064009} {\bibfield
  {journal} {\bibinfo  {journal} {\prd}\ }\textbf {\bibinfo {volume} {80}},\
  \bibinfo {eid} {064009} (\bibinfo {year} {2009}{\natexlab{a}})},\ \Eprint
  {http://arxiv.org/abs/0909.0952} {arXiv:0909.0952} \BibitemShut {NoStop}%
\bibitem [{\citenamefont {{Baumgarte}}\ and\ \citenamefont
  {{Shapiro}}(2009{\natexlab{b}})}]{2009PhRvD..80h9901B}%
  \BibitemOpen
  \bibfield  {author} {\bibinfo {author} {\bibfnamefont {T.~W.}\ \bibnamefont
  {{Baumgarte}}}\ and\ \bibinfo {author} {\bibfnamefont {S.~L.}\ \bibnamefont
  {{Shapiro}}},\ }\href {\doibase 10.1103/PhysRevD.80.089901} {\bibfield
  {journal} {\bibinfo  {journal} {\prd}\ }\textbf {\bibinfo {volume} {80}},\
  \bibinfo {eid} {089901} (\bibinfo {year} {2009}{\natexlab{b}})}\BibitemShut
  {NoStop}%
\bibitem [{\citenamefont {Tichy}(2011)}]{Tichy:2011gw}%
  \BibitemOpen
  \bibfield  {author} {\bibinfo {author} {\bibfnamefont {W.}~\bibnamefont
  {Tichy}},\ }\href {\doibase 10.1103/PhysRevD.84.024041} {\bibfield  {journal}
  {\bibinfo  {journal} {\prd}\ }\textbf {\bibinfo {volume} {84}},\ \bibinfo
  {pages} {024041} (\bibinfo {year} {2011})},\ \Eprint
  {http://arxiv.org/abs/1107.1440} {arXiv:1107.1440 [gr-qc]} \BibitemShut
  {NoStop}%
\bibitem [{\citenamefont {{Tichy}}(2012)}]{2012PhRvD..86f4024T}%
  \BibitemOpen
  \bibfield  {author} {\bibinfo {author} {\bibfnamefont {W.}~\bibnamefont
  {{Tichy}}},\ }\href {\doibase 10.1103/PhysRevD.86.064024} {\bibfield
  {journal} {\bibinfo  {journal} {\prd}\ }\textbf {\bibinfo {volume} {86}},\
  \bibinfo {eid} {064024} (\bibinfo {year} {2012})},\ \Eprint
  {http://arxiv.org/abs/1209.5336} {arXiv:1209.5336} \BibitemShut {NoStop}%
\bibitem [{\citenamefont {{Tsatsin}}\ and\ \citenamefont
  {{Marronetti}}(2013)}]{2013PhRvD..88f4060T}%
  \BibitemOpen
  \bibfield  {author} {\bibinfo {author} {\bibfnamefont {P.}~\bibnamefont
  {{Tsatsin}}}\ and\ \bibinfo {author} {\bibfnamefont {P.}~\bibnamefont
  {{Marronetti}}},\ }\href {\doibase 10.1103/PhysRevD.88.064060} {\bibfield
  {journal} {\bibinfo  {journal} {\prd}\ }\textbf {\bibinfo {volume} {88}},\
  \bibinfo {eid} {064060} (\bibinfo {year} {2013})},\ \Eprint
  {http://arxiv.org/abs/1303.6692} {arXiv:1303.6692} \BibitemShut {NoStop}%
\bibitem [{\citenamefont {Ury{\=u}}\ and\ \citenamefont
  {Tsokaros}(2012)}]{Uryu:2011ky}%
  \BibitemOpen
  \bibfield  {author} {\bibinfo {author} {\bibfnamefont {K.}~\bibnamefont
  {Ury{\=u}}}\ and\ \bibinfo {author} {\bibfnamefont {A.}~\bibnamefont
  {Tsokaros}},\ }\href {\doibase 10.1103/PhysRevD.85.064014} {\bibfield
  {journal} {\bibinfo  {journal} {\prd}\ }\textbf {\bibinfo {volume} {85}},\
  \bibinfo {pages} {064014} (\bibinfo {year} {2012})},\ \Eprint
  {http://arxiv.org/abs/1108.3065} {arXiv:1108.3065 [gr-qc]} \BibitemShut
  {NoStop}%
\bibitem [{\citenamefont {Tsokaros}\ \emph {et~al.}(2015)\citenamefont
  {Tsokaros}, \citenamefont {Ury{\=u}},\ and\ \citenamefont
  {Rezzolla}}]{Tsokaros:2015fea}%
  \BibitemOpen
  \bibfield  {author} {\bibinfo {author} {\bibfnamefont {A.}~\bibnamefont
  {Tsokaros}}, \bibinfo {author} {\bibfnamefont {K.}~\bibnamefont {Ury{\=u}}},
  \ and\ \bibinfo {author} {\bibfnamefont {L.}~\bibnamefont {Rezzolla}},\
  }\href {\doibase 10.1103/PhysRevD.91.104030} {\bibfield  {journal} {\bibinfo
  {journal} {\prd}\ }\textbf {\bibinfo {volume} {91}},\ \bibinfo {pages}
  {104030} (\bibinfo {year} {2015})},\ \Eprint
  {http://arxiv.org/abs/1502.05674} {arXiv:1502.05674 [gr-qc]} \BibitemShut
  {NoStop}%
\bibitem [{\citenamefont {Tsokaros}\ \emph {et~al.}(2016)\citenamefont
  {Tsokaros}, \citenamefont {Mundim}, \citenamefont {Galeazzi}, \citenamefont
  {Rezzolla},\ and\ \citenamefont {Ury{\=u}}}]{Tsokaros:2016eik}%
  \BibitemOpen
  \bibfield  {author} {\bibinfo {author} {\bibfnamefont {A.}~\bibnamefont
  {Tsokaros}}, \bibinfo {author} {\bibfnamefont {B.~C.}\ \bibnamefont
  {Mundim}}, \bibinfo {author} {\bibfnamefont {F.}~\bibnamefont {Galeazzi}},
  \bibinfo {author} {\bibfnamefont {L.}~\bibnamefont {Rezzolla}}, \ and\
  \bibinfo {author} {\bibfnamefont {K.}~\bibnamefont {Ury{\=u}}},\ }\href
  {\doibase 10.1103/PhysRevD.94.044049} {\bibfield  {journal} {\bibinfo
  {journal} {\prd}\ }\textbf {\bibinfo {volume} {94}},\ \bibinfo {pages}
  {044049} (\bibinfo {year} {2016})},\ \Eprint
  {http://arxiv.org/abs/1605.07205} {arXiv:1605.07205 [gr-qc]} \BibitemShut
  {NoStop}%
\bibitem [{\citenamefont {Bernuzzi}\ \emph {et~al.}(2014)\citenamefont
  {Bernuzzi}, \citenamefont {Dietrich}, \citenamefont {Tichy},\ and\
  \citenamefont {Brügmann}}]{Bernuzzi:2013rza}%
  \BibitemOpen
  \bibfield  {author} {\bibinfo {author} {\bibfnamefont {S.}~\bibnamefont
  {Bernuzzi}}, \bibinfo {author} {\bibfnamefont {T.}~\bibnamefont {Dietrich}},
  \bibinfo {author} {\bibfnamefont {W.}~\bibnamefont {Tichy}}, \ and\ \bibinfo
  {author} {\bibfnamefont {B.}~\bibnamefont {Brügmann}},\ }\href {\doibase
  10.1103/PhysRevD.89.104021} {\bibfield  {journal} {\bibinfo  {journal}
  {\prd}\ }\textbf {\bibinfo {volume} {89}},\ \bibinfo {pages} {104021}
  (\bibinfo {year} {2014})},\ \Eprint {http://arxiv.org/abs/1311.4443}
  {arXiv:1311.4443 [gr-qc]} \BibitemShut {NoStop}%
\bibitem [{\citenamefont {Dietrich}\ \emph {et~al.}(2015)\citenamefont
  {Dietrich}, \citenamefont {Moldenhauer}, \citenamefont {Johnson-McDaniel},
  \citenamefont {Bernuzzi}, \citenamefont {Markakis}, \citenamefont
  {Brügmann},\ and\ \citenamefont {Tichy}}]{Dietrich:2015pxa}%
  \BibitemOpen
  \bibfield  {author} {\bibinfo {author} {\bibfnamefont {T.}~\bibnamefont
  {Dietrich}}, \bibinfo {author} {\bibfnamefont {N.}~\bibnamefont
  {Moldenhauer}}, \bibinfo {author} {\bibfnamefont {N.~K.}\ \bibnamefont
  {Johnson-McDaniel}}, \bibinfo {author} {\bibfnamefont {S.}~\bibnamefont
  {Bernuzzi}}, \bibinfo {author} {\bibfnamefont {C.~M.}\ \bibnamefont
  {Markakis}}, \bibinfo {author} {\bibfnamefont {B.}~\bibnamefont {Brügmann}},
  \ and\ \bibinfo {author} {\bibfnamefont {W.}~\bibnamefont {Tichy}},\ }\href
  {\doibase 10.1103/PhysRevD.92.124007} {\bibfield  {journal} {\bibinfo
  {journal} {\prd}\ }\textbf {\bibinfo {volume} {92}},\ \bibinfo {pages}
  {124007} (\bibinfo {year} {2015})},\ \Eprint
  {http://arxiv.org/abs/1507.07100} {arXiv:1507.07100 [gr-qc]} \BibitemShut
  {NoStop}%
\bibitem [{\citenamefont {Tacik}\ \emph {et~al.}(2015)\citenamefont {Tacik}
  \emph {et~al.}}]{Tacik:2015tja}%
  \BibitemOpen
  \bibfield  {author} {\bibinfo {author} {\bibfnamefont {N.}~\bibnamefont
  {Tacik}} \emph {et~al.},\ }\href {\doibase 10.1103/PhysRevD.94.049903,
  10.1103/PhysRevD.92.124012} {\bibfield  {journal} {\bibinfo  {journal}
  {\prd}\ }\textbf {\bibinfo {volume} {92}},\ \bibinfo {pages} {124012}
  (\bibinfo {year} {2015})},\ \bibinfo {note} {[Erratum: Phys. Rev. D 94, no.4,
  049903 (2016)]},\ \Eprint {http://arxiv.org/abs/1508.06986} {arXiv:1508.06986
  [gr-qc]} \BibitemShut {NoStop}%
\bibitem [{\citenamefont {Friedman}\ \emph {et~al.}(2002)\citenamefont
  {Friedman}, \citenamefont {Uryu},\ and\ \citenamefont
  {Shibata}}]{Friedman:2001pf}%
  \BibitemOpen
  \bibfield  {author} {\bibinfo {author} {\bibfnamefont {J.~L.}\ \bibnamefont
  {Friedman}}, \bibinfo {author} {\bibfnamefont {K.}~\bibnamefont {Uryu}}, \
  and\ \bibinfo {author} {\bibfnamefont {M.}~\bibnamefont {Shibata}},\ }\href
  {\doibase 10.1103/PhysRevD.70.129904, 10.1103/PhysRevD.65.064035} {\bibfield
  {journal} {\bibinfo  {journal} {\prd}\ }\textbf {\bibinfo {volume} {65}},\
  \bibinfo {pages} {064035} (\bibinfo {year} {2002})},\ \bibinfo {note}
  {[Erratum: Phys. Rev.D70,129904(2004)]},\ \Eprint
  {http://arxiv.org/abs/gr-qc/0108070} {arXiv:gr-qc/0108070 [gr-qc]}
  \BibitemShut {NoStop}%
\bibitem [{\citenamefont {Friedman}\ and\ \citenamefont
  {Stergioulas}(2013)}]{Friedman:2013xza}%
  \BibitemOpen
  \bibfield  {author} {\bibinfo {author} {\bibfnamefont {J.~L.}\ \bibnamefont
  {Friedman}}\ and\ \bibinfo {author} {\bibfnamefont {N.}~\bibnamefont
  {Stergioulas}},\ }\href {\doibase 10.1017/CBO9780511977596} {\emph {\bibinfo
  {title} {{Rotating Relativistic Stars}}}},\ Cambridge Monographs on
  Mathematical Physics\ (\bibinfo  {publisher} {Cambridge University Press},\
  \bibinfo {year} {2013})\BibitemShut {NoStop}%
\bibitem [{\citenamefont {Ury{\=u}}\ \emph
  {et~al.}(2016{\natexlab{a}})\citenamefont {Ury{\=u}}, \citenamefont
  {Tsokaros}, \citenamefont {Galeazzi}, \citenamefont {Hotta}, \citenamefont
  {Sugimura}, \citenamefont {Taniguchi},\ and\ \citenamefont
  {Yoshida}}]{Uryu:2016dqr}%
  \BibitemOpen
  \bibfield  {author} {\bibinfo {author} {\bibfnamefont {K.}~\bibnamefont
  {Ury{\=u}}}, \bibinfo {author} {\bibfnamefont {A.}~\bibnamefont {Tsokaros}},
  \bibinfo {author} {\bibfnamefont {F.}~\bibnamefont {Galeazzi}}, \bibinfo
  {author} {\bibfnamefont {H.}~\bibnamefont {Hotta}}, \bibinfo {author}
  {\bibfnamefont {M.}~\bibnamefont {Sugimura}}, \bibinfo {author}
  {\bibfnamefont {K.}~\bibnamefont {Taniguchi}}, \ and\ \bibinfo {author}
  {\bibfnamefont {S.}~\bibnamefont {Yoshida}},\ }\href {\doibase
  10.1103/PhysRevD.93.044056} {\bibfield  {journal} {\bibinfo  {journal}
  {\prd}\ }\textbf {\bibinfo {volume} {93}},\ \bibinfo {pages} {044056}
  (\bibinfo {year} {2016}{\natexlab{a}})}\BibitemShut {NoStop}%
\bibitem [{\citenamefont {{Baumgarte}}\ and\ \citenamefont
  {{Shapiro}}(2010)}]{BSBook}%
  \BibitemOpen
  \bibfield  {author} {\bibinfo {author} {\bibfnamefont {T.~W.}\ \bibnamefont
  {{Baumgarte}}}\ and\ \bibinfo {author} {\bibfnamefont {S.~L.}\ \bibnamefont
  {{Shapiro}}},\ }\href@noop {} {\emph {\bibinfo {title} {{Numerical
  Relativity: Solving Einstein's Equations on the Computer}}}}\ (\bibinfo
  {publisher} {Cambridge University Press},\ \bibinfo {year}
  {2010})\BibitemShut {NoStop}%
\bibitem [{\citenamefont {{Rezzolla}}\ and\ \citenamefont
  {{Zanotti}}(2013)}]{Rezzolla_book:2013}%
  \BibitemOpen
  \bibfield  {author} {\bibinfo {author} {\bibfnamefont {L.}~\bibnamefont
  {{Rezzolla}}}\ and\ \bibinfo {author} {\bibfnamefont {O.}~\bibnamefont
  {{Zanotti}}},\ }\href {\doibase 10.1093/acprof:oso/9780198528906.001.0001}
  {\emph {\bibinfo {title} {Relativistic Hydrodynamics}}}\ (\bibinfo
  {publisher} {Oxford University Press},\ \bibinfo {address} {Oxford, UK},\
  \bibinfo {year} {2013})\BibitemShut {NoStop}%
\bibitem [{\citenamefont {{Bonazzola}}\ \emph {et~al.}(1997)\citenamefont
  {{Bonazzola}}, \citenamefont {{Gourgoulhon}},\ and\ \citenamefont
  {{Marck}}}]{1997PhRvD..56.7740B}%
  \BibitemOpen
  \bibfield  {author} {\bibinfo {author} {\bibfnamefont {S.}~\bibnamefont
  {{Bonazzola}}}, \bibinfo {author} {\bibfnamefont {E.}~\bibnamefont
  {{Gourgoulhon}}}, \ and\ \bibinfo {author} {\bibfnamefont {J.-A.}\
  \bibnamefont {{Marck}}},\ }\href {\doibase 10.1103/PhysRevD.56.7740}
  {\bibfield  {journal} {\bibinfo  {journal} {\prd}\ }\textbf {\bibinfo
  {volume} {56}},\ \bibinfo {pages} {7740} (\bibinfo {year} {1997})},\ \Eprint
  {http://arxiv.org/abs/gr-qc/9710031} {arXiv:gr-qc/9710031} \BibitemShut
  {NoStop}%
\bibitem [{\citenamefont {{Asada}}(1998)}]{1998PhRvD..57.7292A}%
  \BibitemOpen
  \bibfield  {author} {\bibinfo {author} {\bibfnamefont {H.}~\bibnamefont
  {{Asada}}},\ }\href {\doibase 10.1103/PhysRevD.57.7292} {\bibfield  {journal}
  {\bibinfo  {journal} {\prd}\ }\textbf {\bibinfo {volume} {57}},\ \bibinfo
  {pages} {7292} (\bibinfo {year} {1998})},\ \Eprint
  {http://arxiv.org/abs/gr-qc/9804003} {arXiv:gr-qc/9804003} \BibitemShut
  {NoStop}%
\bibitem [{\citenamefont {Shibata}(1998)}]{Shibata:1998um}%
  \BibitemOpen
  \bibfield  {author} {\bibinfo {author} {\bibfnamefont {M.}~\bibnamefont
  {Shibata}},\ }\href {\doibase 10.1103/PhysRevD.58.024012} {\bibfield
  {journal} {\bibinfo  {journal} {\prd}\ }\textbf {\bibinfo {volume} {58}},\
  \bibinfo {pages} {024012} (\bibinfo {year} {1998})},\ \Eprint
  {http://arxiv.org/abs/gr-qc/9803085} {arXiv:gr-qc/9803085 [gr-qc]}
  \BibitemShut {NoStop}%
\bibitem [{\citenamefont {Teukolsky}(1998)}]{Teukolsky:1998sh}%
  \BibitemOpen
  \bibfield  {author} {\bibinfo {author} {\bibfnamefont {S.~A.}\ \bibnamefont
  {Teukolsky}},\ }\href {\doibase 10.1086/306082} {\bibfield  {journal}
  {\bibinfo  {journal} {Astrophys. J.}\ }\textbf {\bibinfo {volume} {504}},\
  \bibinfo {pages} {442} (\bibinfo {year} {1998})},\ \Eprint
  {http://arxiv.org/abs/gr-qc/9803082} {arXiv:gr-qc/9803082 [gr-qc]}
  \BibitemShut {NoStop}%
\bibitem [{\citenamefont {Ury{\=u}}\ \emph
  {et~al.}(2016{\natexlab{b}})\citenamefont {Ury{\=u}}, \citenamefont
  {Tsokaros}, \citenamefont {Baiotti}, \citenamefont {Galeazzi}, \citenamefont
  {Sugiyama}, \citenamefont {Taniguchi},\ and\ \citenamefont
  {Yoshida}}]{Uryu:2016pto}%
  \BibitemOpen
  \bibfield  {author} {\bibinfo {author} {\bibfnamefont {K.}~\bibnamefont
  {Ury{\=u}}}, \bibinfo {author} {\bibfnamefont {A.}~\bibnamefont {Tsokaros}},
  \bibinfo {author} {\bibfnamefont {L.}~\bibnamefont {Baiotti}}, \bibinfo
  {author} {\bibfnamefont {F.}~\bibnamefont {Galeazzi}}, \bibinfo {author}
  {\bibfnamefont {N.}~\bibnamefont {Sugiyama}}, \bibinfo {author}
  {\bibfnamefont {K.}~\bibnamefont {Taniguchi}}, \ and\ \bibinfo {author}
  {\bibfnamefont {S.}~\bibnamefont {Yoshida}},\ }\href {\doibase
  10.1103/PhysRevD.94.101302} {\bibfield  {journal} {\bibinfo  {journal}
  {\prd}\ }\textbf {\bibinfo {volume} {94}},\ \bibinfo {pages} {101302}
  (\bibinfo {year} {2016}{\natexlab{b}})},\ \Eprint
  {http://arxiv.org/abs/1606.04604} {arXiv:1606.04604 [astro-ph.HE]}
  \BibitemShut {NoStop}%
\bibitem [{\citenamefont {{Alford}}\ \emph {et~al.}(2005)\citenamefont
  {{Alford}}, \citenamefont {{Braby}}, \citenamefont {{Paris}},\ and\
  \citenamefont {{Reddy}}}]{2005ApJ...629..969A}%
  \BibitemOpen
  \bibfield  {author} {\bibinfo {author} {\bibfnamefont {M.}~\bibnamefont
  {{Alford}}}, \bibinfo {author} {\bibfnamefont {M.}~\bibnamefont {{Braby}}},
  \bibinfo {author} {\bibfnamefont {M.}~\bibnamefont {{Paris}}}, \ and\
  \bibinfo {author} {\bibfnamefont {S.}~\bibnamefont {{Reddy}}},\ }\href
  {\doibase 10.1086/430902} {\bibfield  {journal} {\bibinfo  {journal} {\apj}\
  }\textbf {\bibinfo {volume} {629}},\ \bibinfo {pages} {969} (\bibinfo {year}
  {2005})},\ \Eprint {http://arxiv.org/abs/nucl-th/0411016}
  {arXiv:nucl-th/0411016 [nucl-th]} \BibitemShut {NoStop}%
\bibitem [{\citenamefont {Read}\ \emph {et~al.}(2009)\citenamefont {Read},
  \citenamefont {Lackey}, \citenamefont {Owen},\ and\ \citenamefont
  {Friedman}}]{Read:2008iy}%
  \BibitemOpen
  \bibfield  {author} {\bibinfo {author} {\bibfnamefont {J.~S.}\ \bibnamefont
  {Read}}, \bibinfo {author} {\bibfnamefont {B.~D.}\ \bibnamefont {Lackey}},
  \bibinfo {author} {\bibfnamefont {B.~J.}\ \bibnamefont {Owen}}, \ and\
  \bibinfo {author} {\bibfnamefont {J.~L.}\ \bibnamefont {Friedman}},\ }\href
  {\doibase 10.1103/PhysRevD.79.124032} {\bibfield  {journal} {\bibinfo
  {journal} {\prd}\ }\textbf {\bibinfo {volume} {79}},\ \bibinfo {pages}
  {124032} (\bibinfo {year} {2009})}\BibitemShut {NoStop}%
\bibitem [{\citenamefont {Dietrich}\ \emph {et~al.}(2017)\citenamefont
  {Dietrich}, \citenamefont {Bernuzzi}, \citenamefont {Ujevic},\ and\
  \citenamefont {Tichy}}]{Dietrich:2016lyp}%
  \BibitemOpen
  \bibfield  {author} {\bibinfo {author} {\bibfnamefont {T.}~\bibnamefont
  {Dietrich}}, \bibinfo {author} {\bibfnamefont {S.}~\bibnamefont {Bernuzzi}},
  \bibinfo {author} {\bibfnamefont {M.}~\bibnamefont {Ujevic}}, \ and\ \bibinfo
  {author} {\bibfnamefont {W.}~\bibnamefont {Tichy}},\ }\href {\doibase
  10.1103/PhysRevD.95.044045} {\bibfield  {journal} {\bibinfo  {journal}
  {\prd}\ }\textbf {\bibinfo {volume} {95}},\ \bibinfo {pages} {044045}
  (\bibinfo {year} {2017})},\ \Eprint {http://arxiv.org/abs/1611.07367}
  {arXiv:1611.07367 [gr-qc]} \BibitemShut {NoStop}%
\bibitem [{\citenamefont {{Blanchet}}(2014)}]{2014LRR....17....2B}%
  \BibitemOpen
  \bibfield  {author} {\bibinfo {author} {\bibfnamefont {L.}~\bibnamefont
  {{Blanchet}}},\ }\href {\doibase 10.12942/lrr-2014-2} {\bibfield  {journal}
  {\bibinfo  {journal} {Living Reviews in Relativity}\ }\textbf {\bibinfo
  {volume} {17}},\ \bibinfo {eid} {2} (\bibinfo {year} {2014})},\ \Eprint
  {http://arxiv.org/abs/1310.1528} {arXiv:1310.1528 [gr-qc]} \BibitemShut
  {NoStop}%
\end{thebibliography}%

\end{document}